\def\plotfour#1#2#3#4#5#6#7#8{
\centering \leavevmode \epsfxsize=.45\columnwidth \epsfbox{#1} \hfil \epsfxsize=.45\columnwidth \epsfbox{#2} \\
\centering \leavevmode \epsfxsize=.45\columnwidth \epsfbox{#3} \hfil \epsfxsize=.45\columnwidth \epsfbox{#4} \\
\centering \leavevmode \epsfxsize=.45\columnwidth \epsfbox{#5} \hfil \epsfxsize=.45\columnwidth \epsfbox{#6} \\
\centering \leavevmode \epsfxsize=.45\columnwidth \epsfbox{#7} \hfil \epsfxsize=.45\columnwidth \epsfbox{#8}
}
 
\def\plotthree#1#2#3{
\centering \leavevmode \epsfxsize=.45\columnwidth \epsfbox{#1} \hfil \epsfxsize=.45\columnwidth \epsfbox{#2} \\
\centering \leavevmode \epsfxsize=.45\columnwidth \epsfbox{#3} \hfil
}

\documentstyle[12pt,aaspp4]{article}

\begin{document}

\title{A Reanalysis of the Ultraviolet Extinction from Interstellar Dust in the Large Magellanic Cloud}
\author{K. A. Misselt, Geoffrey C. Clayton \& Karl D. Gordon}
\affil{Department of Physics and Astronomy, Louisiana State University, \\  Baton Rouge, LA 70803-4001 \\ } 
\affil{email: misselt,gclayton,gordon@rouge.phys.lsu.edu} 

\begin{abstract}
We have reanalyzed the Large Magellanic Cloud's (LMC) ultraviolet (UV)
extinction using data from the IUE final archive.
Our new analysis takes advantage of the improved signal--to--noise of
the IUE NEWSIPS reduction, the exclusion of stars with very low
reddening, the careful selection of well matched comparison stars,
and an analysis of the effects of Galactic foreground dust.
Differences between the average extinction curves of
the 30 Dor region and the rest of the LMC are reduced compared to 
previous studies. We find that there is a group of stars with very
weak 2175 \AA\ bumps that lie in or near the region occupied by the
supergiant shell, LMC 2, on the southeast side of 30 Dor.
The average extinction curves inside and outside LMC 2 show a very 
significant difference in 2175 \AA\ bump strength, but their far--UV 
extinctions are similar. While it is unclear whether or not the 
extinction outside the LMC~2 region can be fit with the relation of
Cardelli, Clayton \& Mathis (CCM), sightlines near LMC~2 
cannot be fit with CCM due to their weak 2175 \AA\ bumps.  
While the extinction properties seen in the LMC lie within the range of 
properties seen in the Galaxy, the correlations of UV extinction 
properties with environment seen in the Galaxy do not appear to hold
in the LMC.
\end{abstract}


\section{Introduction}
As our nearest galactic neighbors, the Magellanic Clouds offer a unique opportunity
to study the effects of different galactic environments on dust properties.
Their importance has increased with the recent discovery that the dust in 
starburst galaxies appears to be 
similar to that in the star forming bar of 
the Small Magellanic Cloud (SMC) (Calzetti et. al. 1994; 
Gordon, Calzetti \& Witt 1997, Gordon \& Clayton 1998 [GC]). Understanding
dust extinction properties
in nearby galaxies is a useful tool in interpreting and modeling
observations in a wide range of extragalactic systems. 

Previous studies of the LMC extinction have all arrived at similar conclusions,
e.g. the average LMC extinction curve is characterized by a weaker 2175 \AA\ bump
and a stronger far--UV rise than the average Galactic extinction curve.  
Two early studies (Nandy et. al. 1981;
Koornneef \& Code 1981) found little spatial variation in the LMC extinction
and computed an average LMC extinction curve.  However, both samples were dominated
by stars near the 30  Doradus star forming region and it was thus unclear whether
their average curves applied to the LMC as a whole.  
A study by Clayton \& Martin (1985) expanded the sample to include a larger number
of non--30  Dor stars and reported tentative evidence for differences between the extinction curves
observed in the 30  Dor region and the rest of the LMC.
Fitzpatrick (1985, hereafter F85) expanded the number of available 
reddened stars to 19 including 7 outside of the 30  Dor region, allowing a more
detailed analysis of regional variations.  F85 found a significant difference
between the UV extinction characteristic of the 30  Dor region and that outside
the 30  Dor region.
The average 30  Dor UV extinction curve was found to have a lower
bump strength and stronger far--UV rise 
($\sim$2 units at $\lambda ^{-1} = 7 \mu$m$^{-1}$) than the non--30  Dor stars.
Fitzpatrick (1986, hereafter F86) expanded the sample by 8 lightly reddened stars 
located outside
the 30 Dor region and confirmed the results of F85.
Clayton et al. (1996) measured the extinction of two LMC stars, 
one in 30 Dor and one outside 30 Dor down to $\sim$ 1000 \AA.  
Both extensions seem to be consistent with extrapolations of the 
IUE extinction curves to shorter wavelengths.

As part of a program to quantify the range of extinction behavior in the Local
Group, we have reanalyzed the extinction in the Magellanic Clouds.  
In particular, no analysis has been done since the discovery that the 
UV extinction along most Galactic sightlines could be described by one parameter, 
the ratio of total--to--selective extinction, $R_V=A_V/E(B-V)$
(Cardelli, Clayton, \& Mathis 1989, hereafter CCM).
It is of great interest whether such a relation exists for the Magellanic clouds.
In this paper, we discuss the results for the LMC.  An analysis
of the SMC extinction appears in GC.  

\section{The Data and the Computation of Extinction Curves}
\subsection{The Sample}
Our initial sample of reddened stars consisted of that defined by
F85.  In an effort to expand the sample we searched the
updated electronic catalog of Rousseau et. al. (1978), available via
the SIMBAD database, which consists of $\sim 1800$ LMC stars.  Two initial
cuts of the catalog were made:  (1) stars with spectral types later
than B4 were discarded and (2) we required $B-V\ge 0$.  The first criterion
limits the effects of spectral type mismatches in the resulting
extinction curves, which can be quite large for spectral types later 
than about B3 (e.g. F85).
The second criterion removes unreddened or lightly reddened stars from
consideration.  We note that all the F85 stars were included in the resulting
sample of $\sim 250$ stars while none of the F86 stars were
included as they all had $B-V < 0$.   We then eliminated emission line 
stars and composite--spectrum objects from our list.  The remaining stars
were checked against the IUE database and all of those for which both
long and short wavelength low-dispersion spectra existed (54) were examined in more 
detail.  Only five stars from this sample were found to be both significantly 
reddened and have high S/N IUE spectra.
These stars were added to our sample and their
IUE spectra are listed in Table 1.
We selected 67 unreddened comparison stars from the sample of LMC supergiants
in Fitzpatrick (1988) for use in constructing extinction curves using the pair
method (Massa, Savage \& Fitzpatrick 1983).
Approximate UV spectral types for all of our reddened stars and their
respective comparison stars (see below for a discussion of the selection of 
extinction pairs) were estimated
from a visual comparison of the IUE spectra to the grid of LMC stars with UV spectral
types given in Neubig \& Bruhweiler (1998).  The estimated UV spectral types 
are reported in Table 2.

\begin{deluxetable}{lcc}
\tablewidth{0pt}
\footnotesize
\tablecaption{``New'' Reddened LMC Stars}
\tablehead{
& \colhead{SWP} & \colhead{LWP/LWR} \\
\colhead{SK} & \colhead{Images} & \colhead{Images}
}
\startdata
$-$66  88 & 39129,45383,45384 & LWP18165,23730 \nl
$-$68  23 & 39155 & LWP18198 \nl
$-$69 206 & 36552,39832 & LWP15751 \nl
$-$69 210 & 23270 & LWR17442 \nl
$-$69 279 & 08924 & LWR07672 \nl
\enddata
\end{deluxetable}

\begin{deluxetable}{lcccccccllc}
\tablecaption{Reddened/Unreddened Pairs: Properties}
\tablewidth{0pt}
\footnotesize
%\scriptsize
\tablehead{
& & \multicolumn{6}{c}{Photometry\tablenotemark{a}} & \multicolumn{2}{c}{Spectral Type\tablenotemark{b}} & \\
\colhead{SK} & \colhead{E(B$-$V)$_{Gal}$\tablenotemark{c}} & \colhead{V} & \colhead{B$-$V} & \colhead{U$-$V} & \colhead{J$-$V} & \colhead{H$-$V} & \colhead{K$-$V} & \colhead{Optical} & \colhead{UV} & \colhead{Key\tablenotemark{d}}
}
 
\startdata
$-$66  19 & 0.09: & 12.79 &    0.12 & $-$0.66 & $-$0.35 & $-$0.45 & --      & B4 I    & B0 Ia &  1  \nl
$-$66 169 & 0.03  & 11.56 & $-$0.13 & $-$1.13 & --      & --      & --      & O9.7 Ia & O9 Ia &     \nl
& & & & & & & & & & \nl
$-$66  88 & 0.06  & 12.70 &    0.20 & $-$0.45 & --      & --      & --      & B2:     & B3 Ia &  2 \nl
$-$66 106 & 0.07  & 11.72 & $-$0.08 & $-$0.99 & --      & --      & --      & B2 Ia   & B3 Ia &    \nl
& & & & & & & & & & \nl
$-$67   2 & 0.06  & 11.26 &    0.08 & $-$0.69 & $-$0.18 & $-$0.21 & $-$0.28 & B1.5 Ia & B2 Ia &  3 \nl
$-$66  35 & 0.07: & 11.55 & $-$0.07 & $-$0.95 & --      & --      & --      & B1 Ia   & B2 Ia &    \nl
& & & & & & & & & & \nl
$-$68  23 & 0.06  & 12.81 &    0.22 & $-$0.39 & --      & --      & --      & OB      & B4 Ia &  4 \nl
$-$67  36 & 0.07  & 12.01 & $-$0.08 & $-$0.89 & --      & --      & --      & B2.5 Ia & B3 Ia &    \nl
& & & & & & & & & & \nl
$-$68  26 & 0.04  & 11.67 &    0.13 & $-$0.62 & --      & --      & --      & B8: I   & B3 Ia &  5 \nl
$-$66  35 & 0.07: & 11.55 & $-$0.07 & $-$0.95 & --      & --      & --      & B1 Ia   & B2 Ia &    \nl
& & & & & & & & & & \nl
$-$69 108 & 0.08  & 12.10 &    0.27 & $-$0.22 & $-$0.57 & $-$0.67 & $-$0.75 & B3 I    & B3 Ia &  6 \nl
$-$67  78 & 0.05  & 11.26 & $-$0.04 & $-$0.77 & --      & --      & --      & B3 Ia   & B3 Ia &    \nl
& & & & & & & & & & \nl
$-$70 116 & 0.05  & 12.05 &    0.11 & $-$0.61 & $-$0.37 & $-$0.43 & $-$0.57 & B2 Ia   & B3 Ia &  7 \nl
$-$67 256 & 0.07  & 11.90 & $-$0.08 & $-$0.97 & --      & --      & --      & B1 Ia   & B3 Ia &    \nl
& & & & & & & & & & \nl
& & & & & & & & & & \nl
$-$68 129 & 0.07  & 12.77 &    0.03 & $-$0.81 & --      & --      & --      & B0.5    & O9 Ia &  8 \nl
$-$68  41 & 0.05  & 12.0  & $-$0.14 & $-$1.10 & --      & --      & --      & B0.5 Ia & B0 Ia &    \nl
& & & & & & & & & & \nl
$-$68 140 & 0.04  & 12.72 &    0.06 & $-$0.77 & $-$0.26 & $-$0.31 & $-$0.37 & B0:     & B0 Ia &  9 \nl
$-$68  41 & 0.05  & 12.0  & $-$0.14 & $-$1.10 & --      & --      & --      & B0.5 Ia & B0 Ia &    \nl
& & & & & & & & & & \nl
$-$68 155 & 0.02  & 12.72 &    0.03 & $-$0.79 & --      & --      & --      & B0.5    & O8 Ia & 10 \nl
$-$67 168 & 0.03  & 12.08 & $-$0.17 & $-$1.17 & --      & --      & --      & O8 Iaf  & O8 Ia &    \nl
& & & & & & & & & & \nl
$-$69 206 & 0.08  & 12.84 &    0.14 & $-$0.62 & --      & --      & --      & B2:     & O9 Ia & 11 \nl
$-$67   5 & 0.06  & 11.34 & $-$0.12 & $-$1.07 & --      & --      & --      & O9.7 Ib & B0 Ia &    \nl
& & & & & & & & & & \nl
$-$69 210 & 0.07  & 12.59 &    0.36 & $-$0.23 & --      & --      & --      & B1.5:   & B1 Ia & 12 \nl
$-$66 118 & 0.08  & 11.81 & $-$0.05 & $-$0.91 & --      & --      & --      & B2 Ia   & B3 Ia &    \nl
& & & & & & & & & & \nl
$-$69 213 & 0.08  & 11.97 &    0.10 & $-$0.65 & $-$0.26 & $-$0.29 & $-$0.33 & B1      & B1 Ia & 13 \nl
$-$70 120 & 0.06  & 11.59 & $-$0.06 & $-$0.94 &    0.21 &    0.25 &    0.14 & B1 Ia   & B1.5 Ia &  \nl
& & & & & & & & & & \nl
$-$69 228 & 0.06  & 12.12 &    0.07 & $-$0.69 & $-$0.10 & $-$0.14 & $-$0.14 & OB      & B2 Ia & 14 \nl
$-$65  15 & 0.12  & 12.14 & $-$0.10 & $-$1.02 & --      & --      & --      & B1 Ia   & B1 Ia &    \nl
& & & & & & & & & & \nl
$-$69 256 & 0.07  & 12.61 &    0.03 & $-$0.80 &    0.03 &    0.04 & $-$0.02 & B0.5    & B1 Ia & 15 \nl
$-$68  41 & 0.05  & 12.0  & $-$0.14 & $-$1.00 & --      & --      & --      & B0.5 Ia & B0 Ia &    \nl
& & & & & & & & & & \nl
$-$69 265 & 0.06  & 11.88 &    0.12 & $-$0.51 & --      & --      & --      & B3 I    & B3 Ia & 16 \nl
$-$68  40 & 0.05  & 11.71 & $-$0.07 & $-$0.86 & --      & --      & --      & B2.5 Ia & B3 Ia &    \nl
& & & & & & & & & & \nl
$-$69 270 & 0.05  & 11.27 &    0.14 & $-$0.52 & $-$0.32 & $-$0.40 & $-$0.46 & B3 Ia   & B2 Ia & 17 \nl
$-$67 228 & 0.03  & 11.49 & $-$0.05 & $-$0.87 & --      & --      & --      & B2 Ia   & B2 Ia &    \nl
& & & & & & & & & & \nl
$-$69 279 & 0.02  & 12.79 &    0.05 & $-$0.79 & $-$0.19 & $-$0.28 & $-$0.34 & OB0     & O9 Ia & 18 \nl
$-$65  63 & 0.03  & 12.56 & $-$0.16 & $-$1.18 & --      & --      & --      & O9.7 I: & O9 Ia &    \nl
& & & & & & & & & & \nl
$-$69 280 & 0.05  & 12.66 &    0.09 & $-$0.65 & $-$0.22 & $-$0.22 & $-$0.33 & B1      & B1.5 Ia & 19 \nl
$-$67 100 & 0.05  & 11.95 & $-$0.09 & $-$0.95 & --      & --      & --      & B1 Ia   & B1 Ia &      \nl
& & & & & & & & & & \nl
\enddata
  
\tablenotetext{a}{Optical photometry from Rousseau et. al. (1978), F85 and Fitzpatrick (1988). IR photometry from Morgan \& Nandy (1982)
and Clayton \& Martin (1985).}
\tablenotetext{b}{Optical spectral types from Rousseau et. al. (1978), F85 and Fitzpatrick (1988). UV spectral types estimated by comparison with LMC UV spectral types of Neubig \& Bruhweiler (1998).}
\tablenotetext{c}{Galactic foreground reddening from Oestreicher et. al. (1995); Colon designates uncertain value.}
\tablenotetext{d}{Key to position in Figure~\ref{fig_ha_map}.}
  
\end{deluxetable}

An implicit assumption of the pair method
is that the Galactic 
foreground reddening is the same
for both the program and comparison stars and, hence, cancels out of the
resulting LMC extinction curve.  As pointed out by several 
authors (e.g. Schwering \& Israel 1991; Oestreicher, Gochermann, \& Schmidt--Kaler 1995), the Galactic
foreground towards the LMC is quite variable ranging from $E(B-V)_{Gal}=0.00$ to 0.17. 
Schwering \& Israel (1991) constructed a foreground reddening map towards the LMC using HI data
and a relationship between $E(B-V)$ and the HI column density. They examined the
F85 and F86 stars at a spatial resolution of 48\arcmin\ (the resolution
of the HI data) and found systematically higher
Galactic foreground reddening associated with the comparison stars than the reddened stars. Accounting for 
this systematic affect reduced the difference between the 30  Dor and non--30  Dor
extinction curves.  
Oestreicher et. al. (1995) used reddenings to $\sim1400$ LMC foreground stars to construct
a Galactic foreground reddening map with a resolution of $\sim$10\arcmin.
We have quantified the differences in Galactic foreground
reddening for our sample using the higher resolution
map of Oestreicher et. al. (1995).  For all but one of our pairs in the 30 Dor
sample, the difference in the Galactic foreground reddening between the 
reddened and comparison stars, $|\Delta$E(B$-$V)$_{Gal}| \le 0.02$ 
while for the non--30  Dor sample,
$|\Delta$E(B$-$V)$_{Gal}| \le 0.03$ for all but one pair as well.  There is 
no systematic difference in the foreground reddening between program and comparison
stars in either sub--sample with the average
$\Delta$E(B$-$V)$_{Gal}$ being near 0 for both samples.  The values for the
Galactic foreground component of the reddening for each star used in the analysis
is given in Table 2.
For the two pairs with large
foreground differences (SK $-$66 19/SK $-$66 169, SK $-$69 228/SK $-$65 15) we 
have estimated the maximum effect on the extinction  curve
to be less than the photometric uncertainties.
Therefore, we have not corrected the individual 
curves for the differences in the Galactic foreground.

\subsection{The Extinction Curves}
Extinction curves were constructed using the standard pair method (e.g. Massa, Savage
\& Fitzpatrick 1983).
Short and long wavelength $IUE$ spectra were extracted using the
$IUE$ NEWSIPS reduction, co--added,
binned to the instrumental resolution of $\sim$5~\AA~ and merged at
the maximum wavelength in the short wavelength spectrum.  
Uncertainties in the extinction curve contain terms that depend both
on the broadband photometric uncertainties as well as uncertainties in the
$IUE$ fluxes.  The flux uncertainties are now calculated 
directly in the NEWSIPS reduction.
For details of our error analysis, the reader is referred
to GC.

\begin{figure}[tbh]
\begin{center}
\plotone{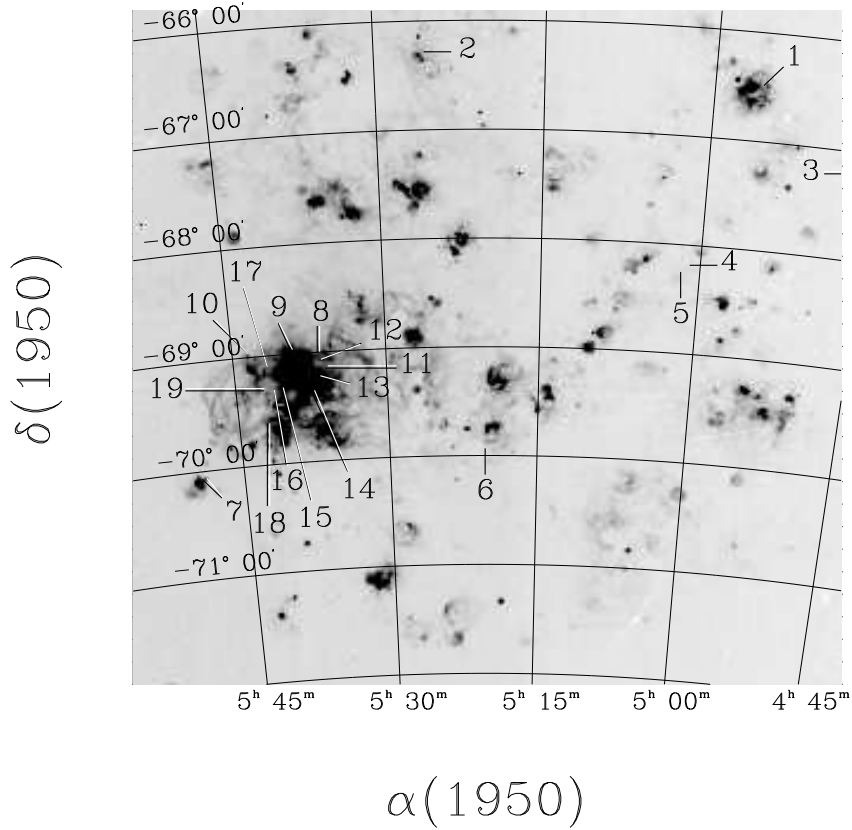}
\caption{Positions of reddened stars plotted on an H$\alpha$ image. A key to the numbering is
provided in Table 2. \label{fig_ha_map} }
\end{center}
\end{figure}

Previous studies suffered
from systematic temperature and luminosity mismatches between the unreddened/reddened
star pairs.  These mismatches were evident in the imperfect line cancellations seen
in the extinction curves, especially the Fe~III blend near 5.1~$\micron^{-1}$.  
This study minimizes
mismatches by using a larger sample of comparison stars
than was available to previous studies.
Comparison stars for each reddened star were selected to satisfy the three Fitzpatrick
criteria (F85); in addition, we required $\Delta (B-V) \ge 0.15$ between the 
reddened and comparison stars to minimize the uncertainties in the extinction curve.  
The first
criterion requires that $\Delta (U-B)/\Delta (B-V)$ be appropriate to dust reddening.
The average value of $\Delta (U-B)/\Delta (B-V)$ for the LMC
is $0.83\pm0.1$ (F85).  Stars with $0.63 \le \Delta (U-B)/\Delta (B-V) \le 1.03$ were selected. 
The second criterion requires that the difference in intrinsic $V$
magnitudes between the comparison and reddened stars be ``small'' ($\mid \Delta V \mid < 0.8$).
The $V$ magnitudes of our program stars were dereddened 
assuming $R_V = 3.1$.
As all LMC stars are at roughly the same distance, this criterion amounts to assuring comparable
absolute magnitudes between the comparison and reddened stars thus minimizing luminosity
mismatches.    The third criterion
requires that the comparison and reddened star UV spectra 
be well--matched.  This minimizes residual features in the extinction curve
not due to extinction.
This procedure resulted in 3--10
potential comparison stars for each reddened star.  Each potential comparison star
was used to compute an extinction curve. The reddened/comparison star pair
which resulted in a curve with the smallest line residuals was adopted.
Five stars from the F85 sample had $\Delta (B-V) < 0.15$ and were discarded,
leaving a total of 19 reddened stars in our study.
These included three 30 Dor stars (SK 
$-$68 126, $-$69 199 and $-$69 282)  and two non--30 Dor stars (SK $-$68 107 and $-$71 52). 
In addition, five stars have been added, three to the 30 Dor sample (SK $-$69 206, 
$-$69 210 and 
$-$69 279)  and two to the non--30 Dor sample (SK $-$66 88 and $-$68 23). 
We have indicated the positions of all of our stars
on an H$\alpha$ map of the LMC in Figure~\ref{fig_ha_map}.  A key to the numbering of 
the stellar positions
in Figure~\ref{fig_ha_map} is given in Table 2.

\begin{figure}
\begin{center}
        \plotthree{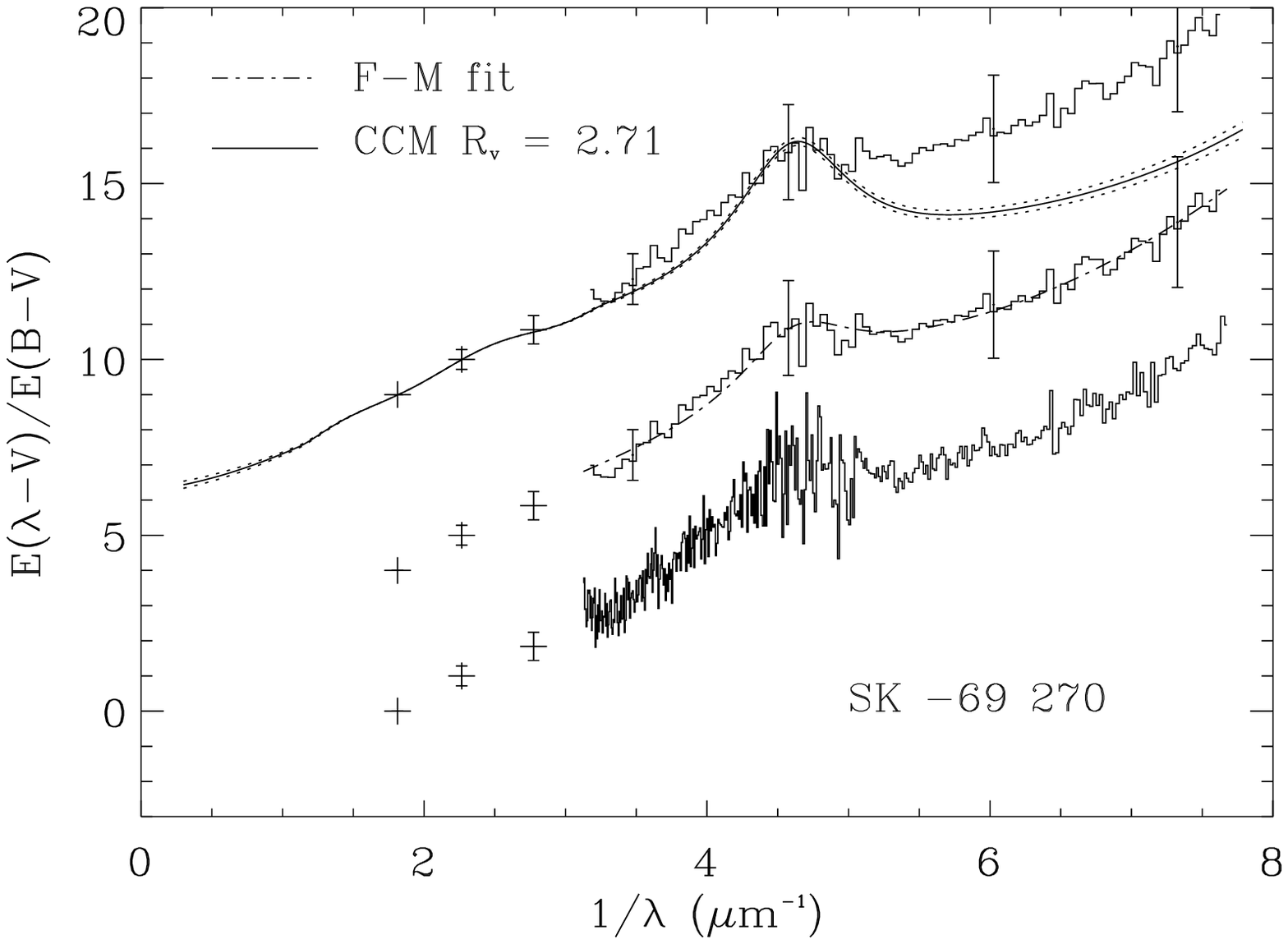}{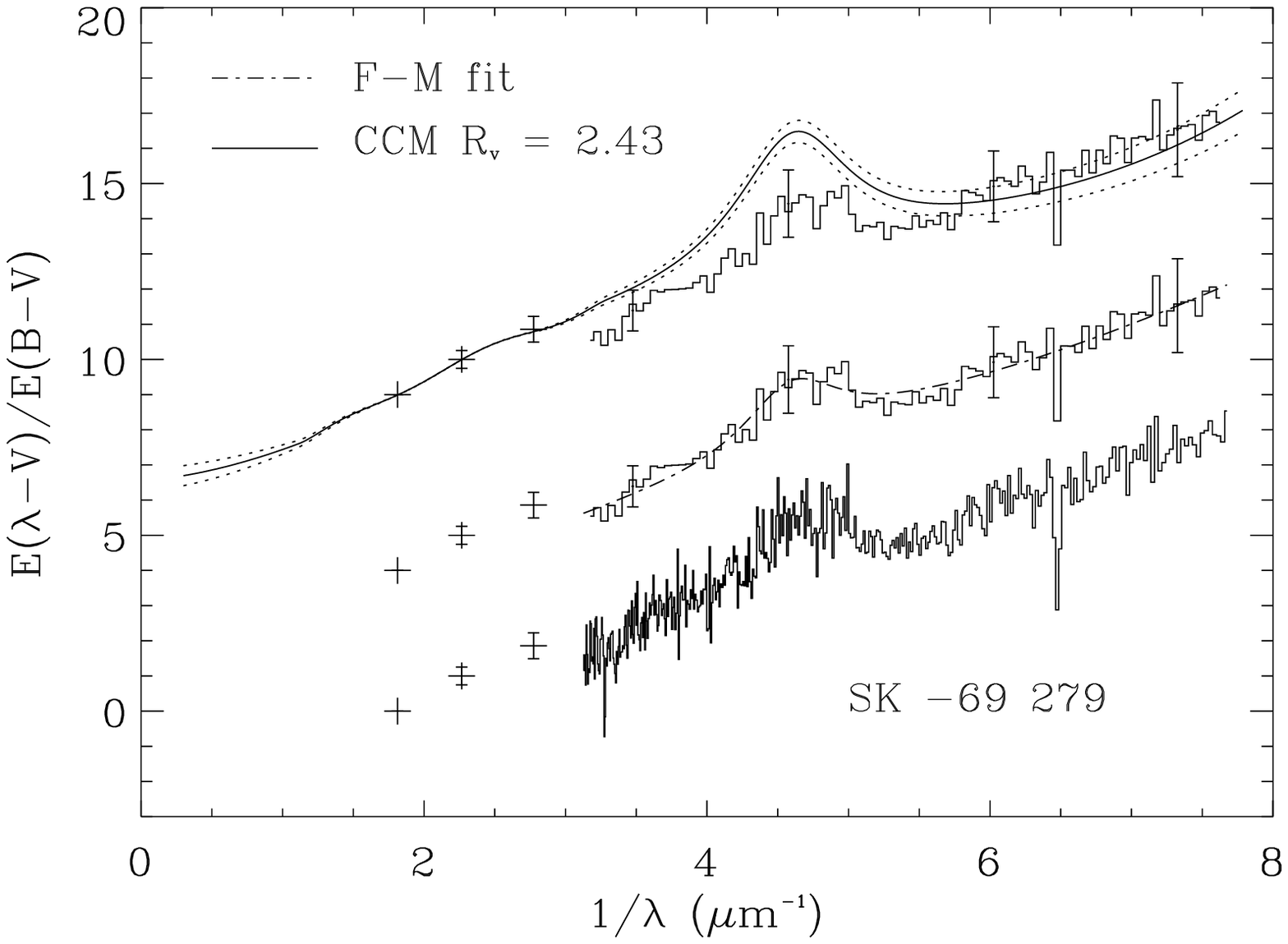}
              {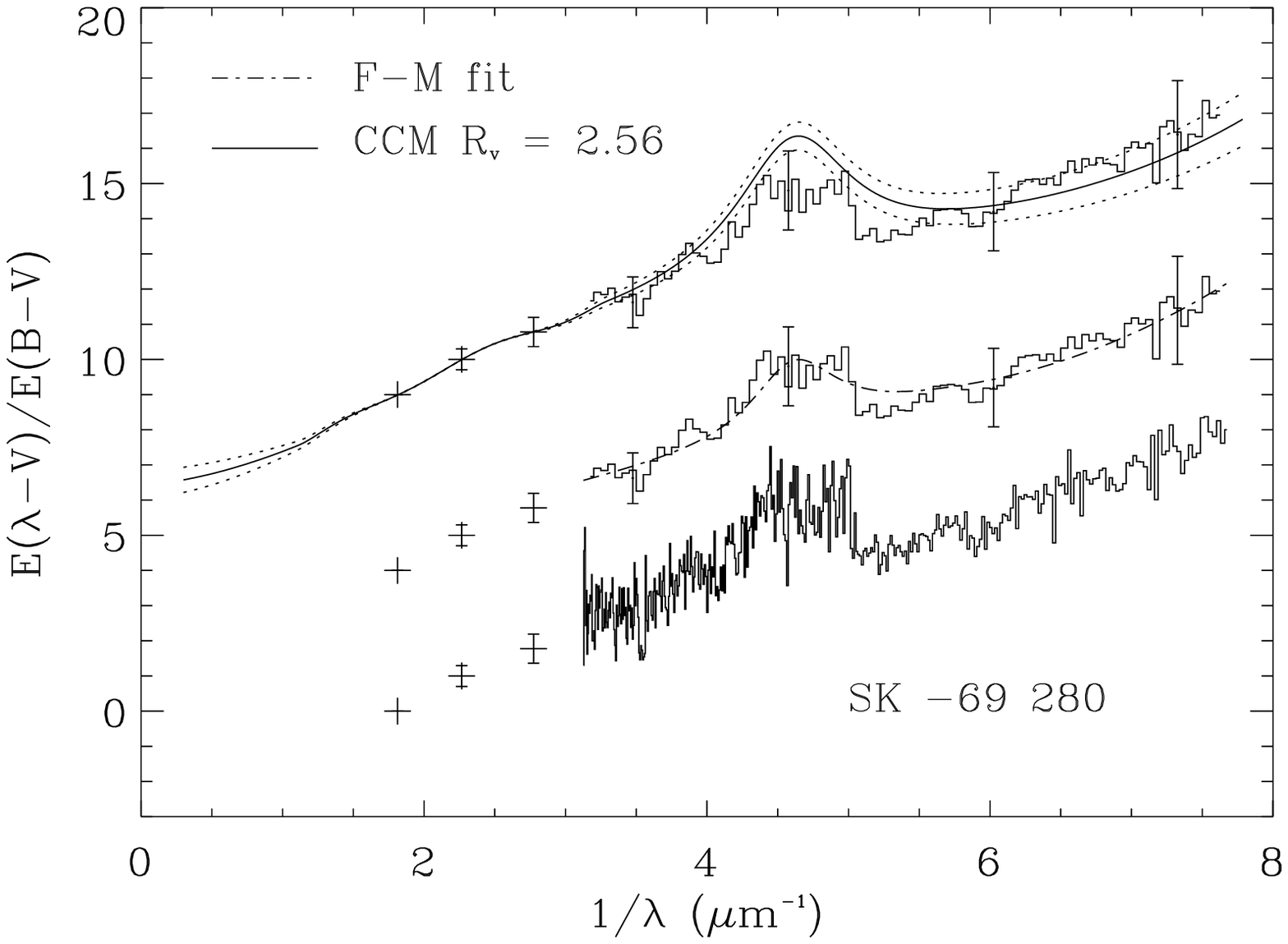}
\caption{Individual LMC extinction curves.  Optical data are included and,
when available, IR. We have plotted the FM fits and CCM curves
offset by 4 and 9 units, respectively along with the re--binned
extinction curve. Where measured values of R$_V$ are available, CCM
curves for the measured R$_V$ (solid line) and R$_V\pm \sigma_{R_V}$
are plotted (dotted line).  When no measured value of R$_V$ was
available, the ``best fit'' CCM curve is plotted.  If no single
value of R$_V$ provided an adequate fit, a CCM curve with
R$_V = 3.1$ is plotted. \label{fig_ext_curves} }
\end{center}
\end{figure}

The final extinction
curves computed for each pair are shown in Figure~\ref{fig_ext_curves} and the 
star pairs are listed in Table 2.  The extinction curves have been
fit using the Fitzpatrick \& Massa (1990, hereafter FM) parameterization. 
The FM fit is a six parameter fit including a linear background, a Drude profile representing
the 2175 \AA\ bump, and a far--UV curvature term.  We emphasize that this parameterization is
empirical and the individual functions describing the extinction curve probably have
limited physical significance (Mathis \& Cardelli 1992).
The FM fits to individual extinction curves are plotted 
in Figure~\ref{fig_ext_curves} and the best fit parameters for each curve are
given in Table 3; the functional form of the parameterization is given as a footnote to 
Table 3.  
In determining the uncertainties on the individual fit parameters we have considered the effects
of two sources of uncertainty, photometric and spectral mismatch. The photometric uncertainties
include those in the broad band optical photometry as well as those in the IUE
fluxes (for a detailed discussion of these uncertainties, see GC).  
We estimate their effect on the FM parameters by shifting the extinction
curves upward by $1\sigma$ and downward by $1\sigma$ point--by--point.  FM fits were
made to both of the shifted extinction curves and the error in each individual parameter
is taken as one--half the absolute value of the difference in the fit parameters 
between the two curves.  The photometric uncertainties contribute most significantly to 
errors in the FM parameters $C_1, C_2, C_3$, and $C_4$; they have little effect on the
bump parameters $x_0$ and $\gamma$.  
The effects of mismatch errors on the FM parameters were taken from Cardelli et. al.
(1992).
By varying the spectral type of the comparison star and fitting the resulting extinction
curve, they were able to estimate the uncertainties introduced in the FM fit parameters (Table
6 of Cardelli et. al. 1992).  We adopt the quadrature sum of the these two sources of uncertainty
as our estimate of the uncertainties in the individual FM fit parameters (Table 3).
For weak features (ie. the weak bump lines of sight in our sample), 
the uncertainties introduced by spectral mismatches may be underestimated by the
adopted uncertainties.

\begin{deluxetable}{lcccccccc}
\tablewidth{0pt}
\scriptsize
\tablecaption{FM Fit Parameters}
 
\tablehead{
& & \multicolumn{6}{c}{FM Fit Parameters\tablenotemark{a}} & \\
\colhead{SK} & \colhead{$\Delta$(B$-$V)} & \colhead{$x_{0}$} & \colhead{$\gamma$} & \colhead{$C_1$} & \colhead{$C_2$} & \colhead{$C_3$} & \colhead{$C_4$} & \colhead{$C_3/\gamma^2$}
}
 
\startdata
& \multicolumn{7}{c}{LMC--Average Sample} & \\
\cline{1-9}
& & & & & & & & \nl
$-$66  19 & 0.25 & 4.653$\pm$0.010 & 0.97$\pm$0.07 & $+$0.09$\pm$0.44 & 0.75$\pm$0.11 & 2.34$\pm$0.42 & 0.91$\pm$0.12 & 2.49$\pm$0.57 \nl
$-$66  88 & 0.28 & 4.579$\pm$0.019 & 1.03$\pm$0.06 & $-$0.88$\pm$0.38 & 1.00$\pm$0.13 & 2.77$\pm$0.46 & 0.48$\pm$0.10 & 2.61$\pm$0.53 \nl
$-$67   2 & 0.15 & 4.625$\pm$0.010 & 1.08$\pm$0.07 & $-$3.59$\pm$0.40 & 1.67$\pm$0.26 & 3.71$\pm$0.46 & 0.91$\pm$0.20 & 3.18$\pm$0.57 \nl
$-$68  23 & 0.30 & 4.513$\pm$0.037 & 1.05$\pm$0.06 & $+$0.11$\pm$0.42 & 0.65$\pm$0.10 & 4.28$\pm$0.84 & 0.71$\pm$0.14 & 3.88$\pm$0.88 \nl
$-$68  26 & 0.20 & 4.671$\pm$0.012 & 1.10$\pm$0.06 & $-$0.64$\pm$0.43 & 0.90$\pm$0.13 & 3.76$\pm$0.44 & 0.43$\pm$0.11 & 3.11$\pm$0.50 \nl
$-$68 129 & 0.17 & 4.587$\pm$0.011 & 0.73$\pm$0.06 & $-$1.48$\pm$0.39 & 1.26$\pm$0.19 & 1.50$\pm$0.42 & 0.72$\pm$0.16 & 2.81$\pm$0.91 \nl
$-$69 108 & 0.31 & 4.574$\pm$0.011 & 1.04$\pm$0.06 & $-$1.25$\pm$0.39 & 0.98$\pm$0.11 & 4.31$\pm$0.44 & 0.54$\pm$0.10 & 3.98$\pm$0.61 \nl
$-$69 206 & 0.26 & 4.519$\pm$0.034 & 0.65$\pm$0.05 & $-$1.40$\pm$0.38 & 1.23$\pm$0.14 & 1.08$\pm$0.43 & 0.38$\pm$0.11 & 2.56$\pm$1.09 \nl
$-$69 210 & 0.41 & 4.669$\pm$0.011 & 0.67$\pm$0.06 & $-$1.15$\pm$0.37 & 1.12$\pm$0.11 & 1.42$\pm$0.41 & 0.52$\pm$0.11 & 3.16$\pm$1.07 \nl
$-$69 213 & 0.16 & 4.570$\pm$0.017 & 0.77$\pm$0.05 & $-$2.62$\pm$0.37 & 1.56$\pm$0.24 & 2.08$\pm$0.46 & 0.83$\pm$0.21 & 3.51$\pm$0.90 \nl
& & & & & & & & \nl
Average\tablenotemark{b} & 0.25 & 4.596$\pm$0.017 & 0.91$\pm$0.05 & $-$1.28$\pm$0.34 & 1.11$\pm$0.10 & 2.73$\pm$0.37 & 0.64$\pm$0.06 & 3.13$\
pm$0.16 \nl
& & & & & & & & \nl
& \multicolumn{7}{c}{LMC 2 Sample} & \nl
\cline{1-9}
& & & & & & & & \nl
$-$68 140 & 0.20 & 4.559$\pm$0.022 & 1.07$\pm$0.09 & $-$1.02$\pm$0.40 & 1.13$\pm$0.17 & 1.62$\pm$0.41 & 0.77$\pm$0.14 & 1.41$\pm$0.43 \nl
$-$68 155 & 0.20 & 4.663$\pm$0.011 & 0.91$\pm$0.07 & $-$4.38$\pm$0.42 & 1.82$\pm$0.23 & 2.06$\pm$0.41 & 0.30$\pm$0.11 & 2.49$\pm$0.62 \nl
$-$69 228 & 0.17 & 4.658$\pm$0.016 & 1.26$\pm$0.13 & $-$2.33$\pm$0.38 & 1.20$\pm$0.18 & 2.30$\pm$0.42 & 0.17$\pm$0.09 & 1.45$\pm$0.40 \nl
$-$69 256 & 0.17 & 4.622$\pm$0.038 & 1.21$\pm$0.05 & $-$2.50$\pm$0.37 & 1.30$\pm$0.19 & 2.15$\pm$0.51 & 0.30$\pm$0.11 & 1.47$\pm$0.37 \nl
$-$69 265 & 0.19 & 4.627$\pm$0.018 & 0.92$\pm$0.10 & $-$2.47$\pm$0.39 & 1.37$\pm$0.18 & 0.88$\pm$0.41 & 0.18$\pm$0.11 & 1.04$\pm$0.54 \nl
$-$69 270 & 0.19 & 4.651$\pm$0.011 & 1.12$\pm$0.09 & $-$2.26$\pm$0.37 & 1.53$\pm$0.21 & 2.66$\pm$0.45 & 0.74$\pm$0.15 & 2.12$\pm$0.49 \nl
$-$69 279 & 0.21 & 4.603$\pm$0.016 & 0.84$\pm$0.06 & $-$2.73$\pm$0.37 & 1.36$\pm$0.16 & 1.33$\pm$0.42 & 0.17$\pm$0.10 & 1.88$\pm$0.65 \nl
$-$69 280 & 0.18 & 4.618$\pm$0.016 & 0.74$\pm$0.06 & $-$0.51$\pm$0.49 & 0.96$\pm$0.14 & 1.15$\pm$0.41 & 0.64$\pm$0.15 & 2.10$\pm$0.82 \nl
$-$70 116 & 0.19 & 4.637$\pm$0.024 & 1.42$\pm$0.10 & $-$1.22$\pm$0.45 & 1.09$\pm$0.15 & 3.13$\pm$0.41 & 0.54$\pm$0.13 & 1.55$\pm$0.30 \nl
& & & & & & & & \nl
Average\tablenotemark{b} & 0.19 & 4.626$\pm$0.010 & 1.05$\pm$0.07 & $-$2.16$\pm$0.36 & 1.31$\pm$0.08 & 1.92$\pm$0.23 & 0.42$\pm$0.08 & 1.72$\
pm$0.14 \nl
& & & & & & & & \nl
& \multicolumn{7}{c}{Milky Way average\tablenotemark{c}} & \nl
\cline{1-9}
& & & & & & & & \nl
& -- & 4.596$\pm$0.002 & 0.96$\pm$0.01 & 0.12$\pm$0.11 & 0.63$\pm$0.04 & 3.26$\pm$0.11 & 0.41$\pm$0.02 & 3.49$\pm$0.07 \nl
\enddata
\tablenotetext{a}{Analytic fit to extinction curve following FM: \nl
                  \begin{center}
                  $\frac{\Delta(\lambda-V)}{\Delta(B-V)}=C_1+C_2x+C_3D(x)+C_4F(x),$
                  \end{center}
                  where $x=\lambda^{-1}$, and  \nl
                  \begin{center}
                  $D(x) = \frac{x^2}{(x^2-x_0^2)^2+x^2\gamma^2}.$
                  \end{center}
                  \begin{center}
                  $F(x) = 0.5329(x-5.9)^2+0.05644(x-5.9)^3~~~(x > 5.9)$
                  \end{center}
                  and $F(x) = 0$ otherwise.
                  }
\tablenotetext{b}{Uncertainties in the averages quoted as the standard deviation of the sample mean for the
                  respective samples, eg. $\sigma _{i}/\sqrt{N}$.}
\tablenotetext{c}{From the Galactic data of FM.  Errors are the standard deviation of the sample mean.}
\end{deluxetable}

\begin{figure}[tbp]
\begin{center}
        \plotfour{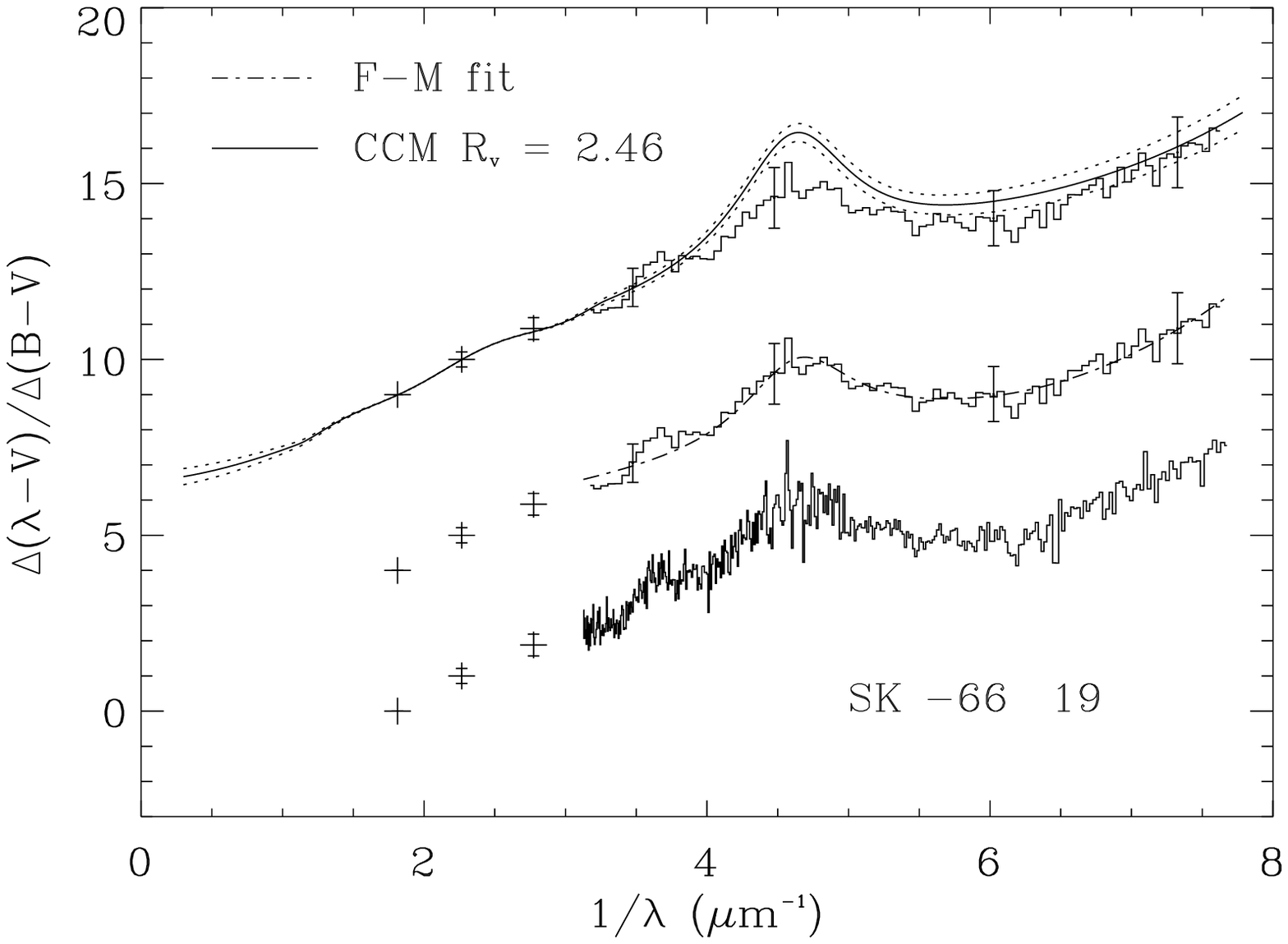}{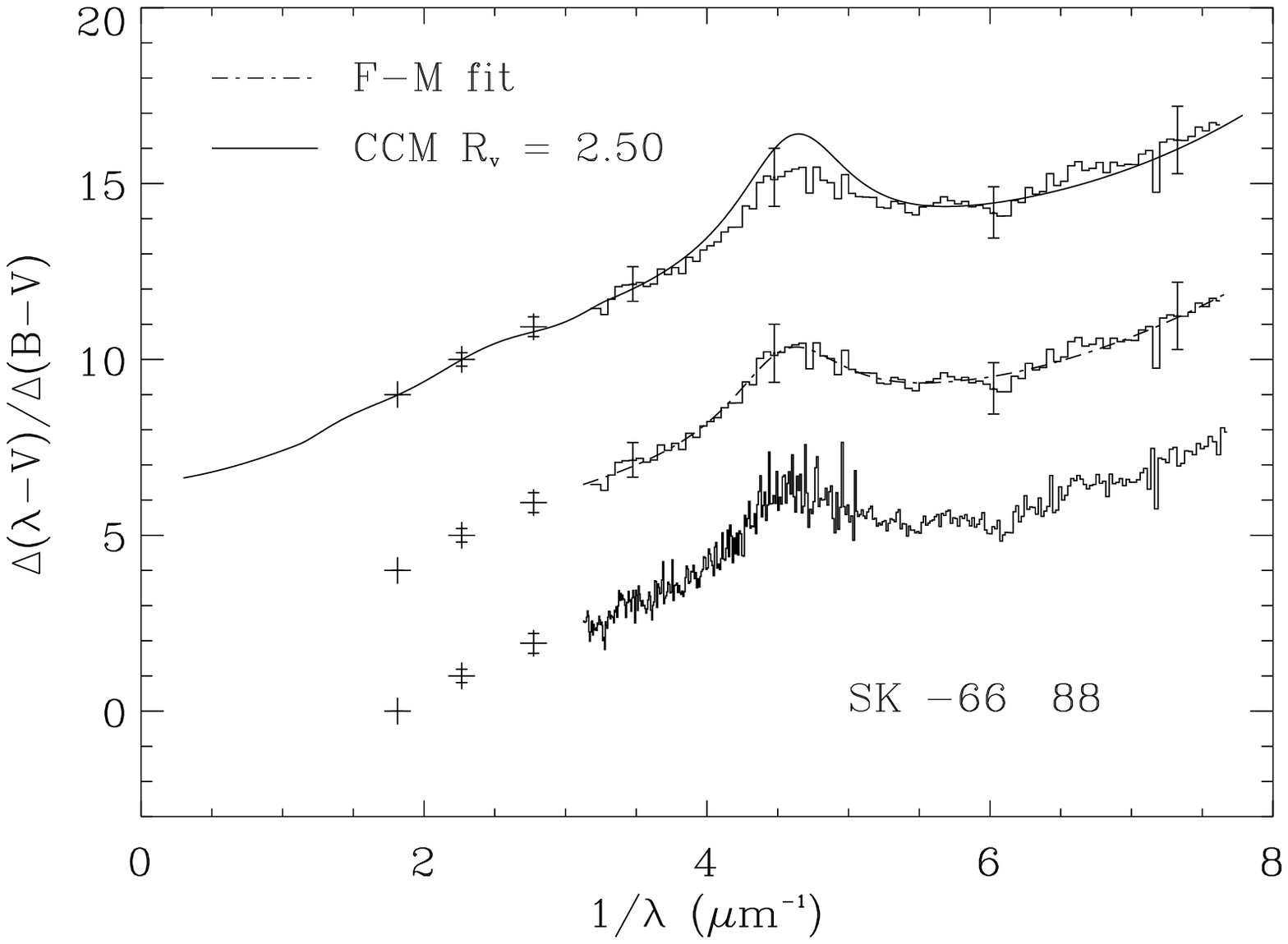}
                 {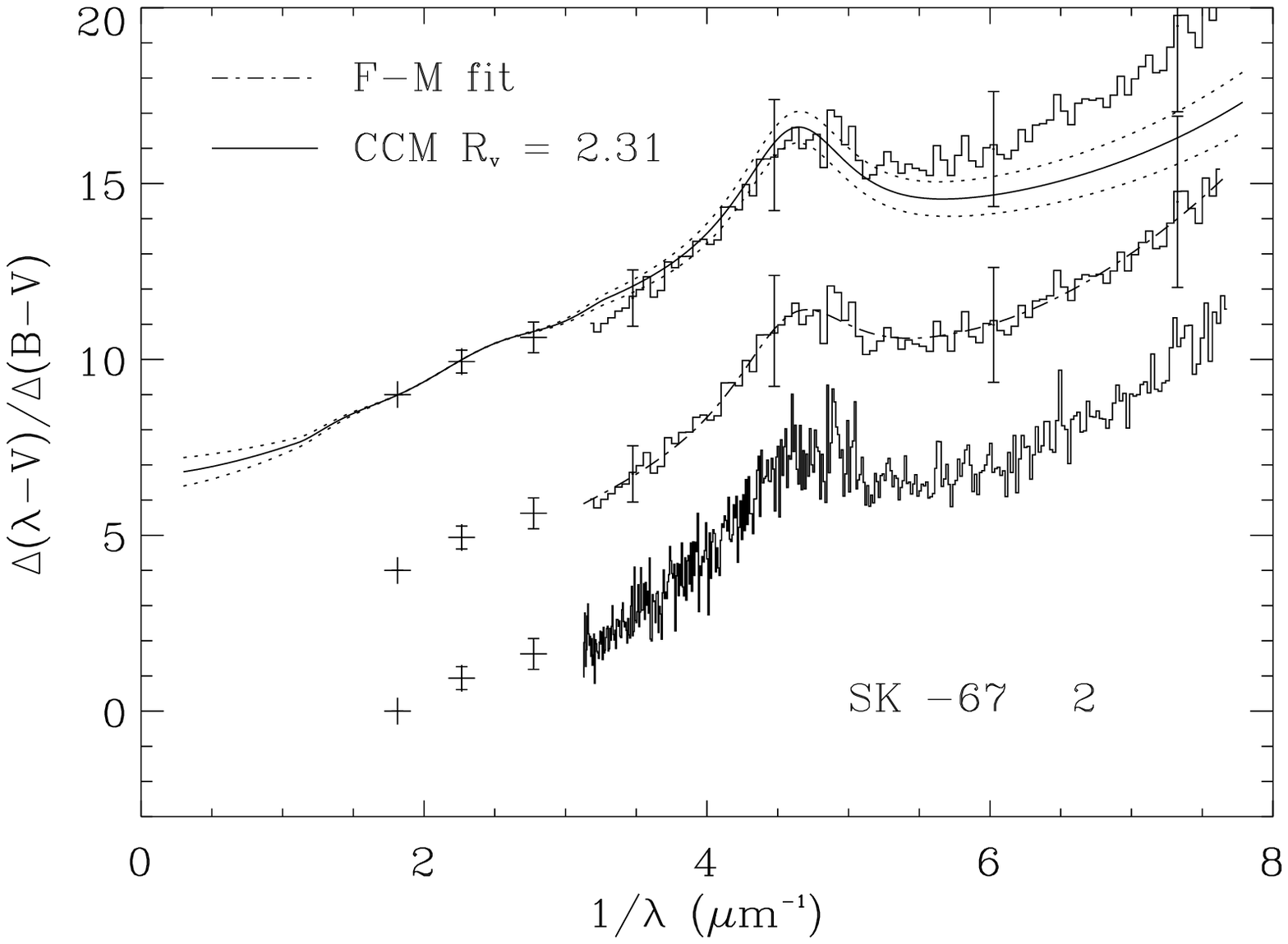}{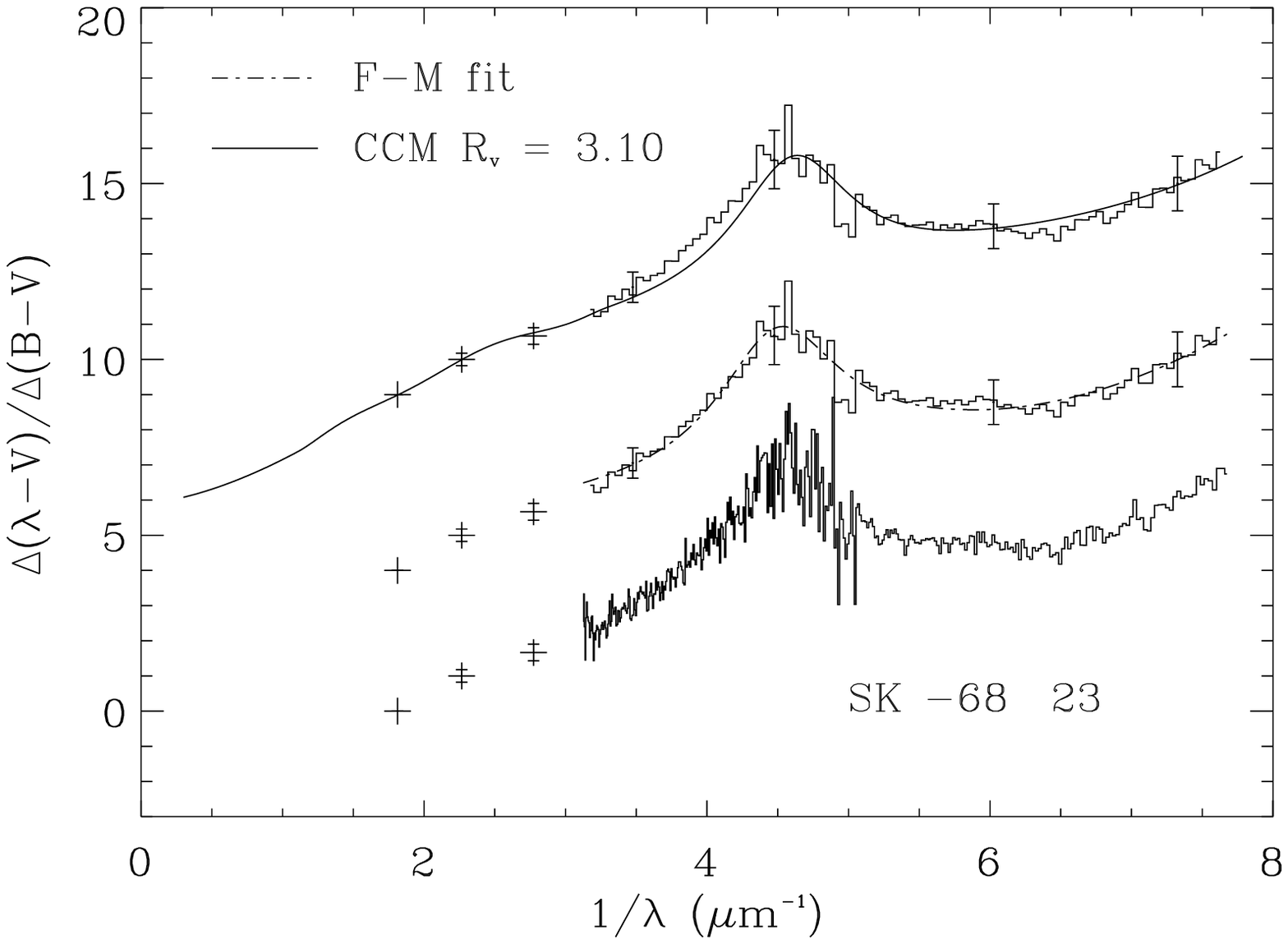}
                 {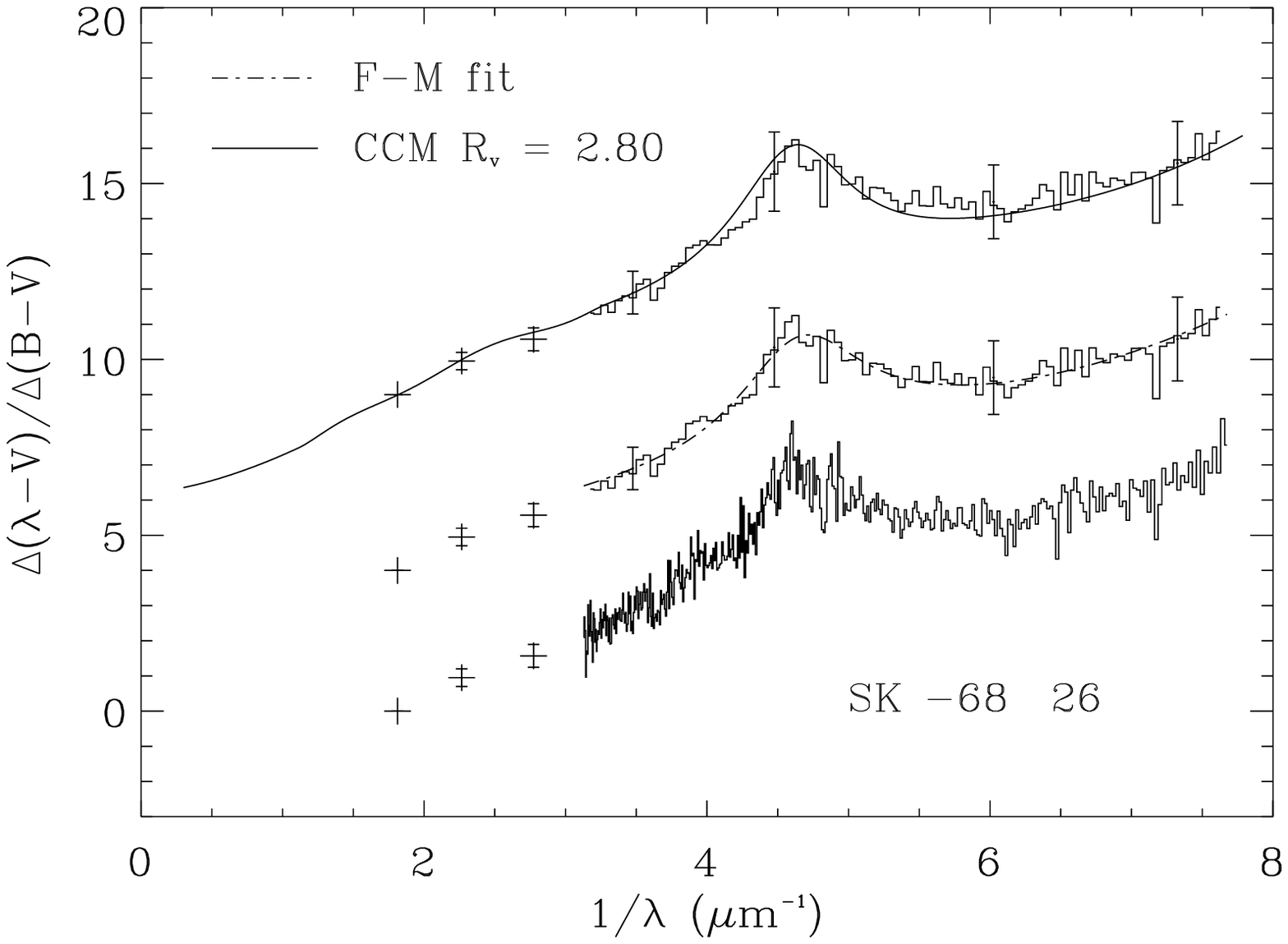}{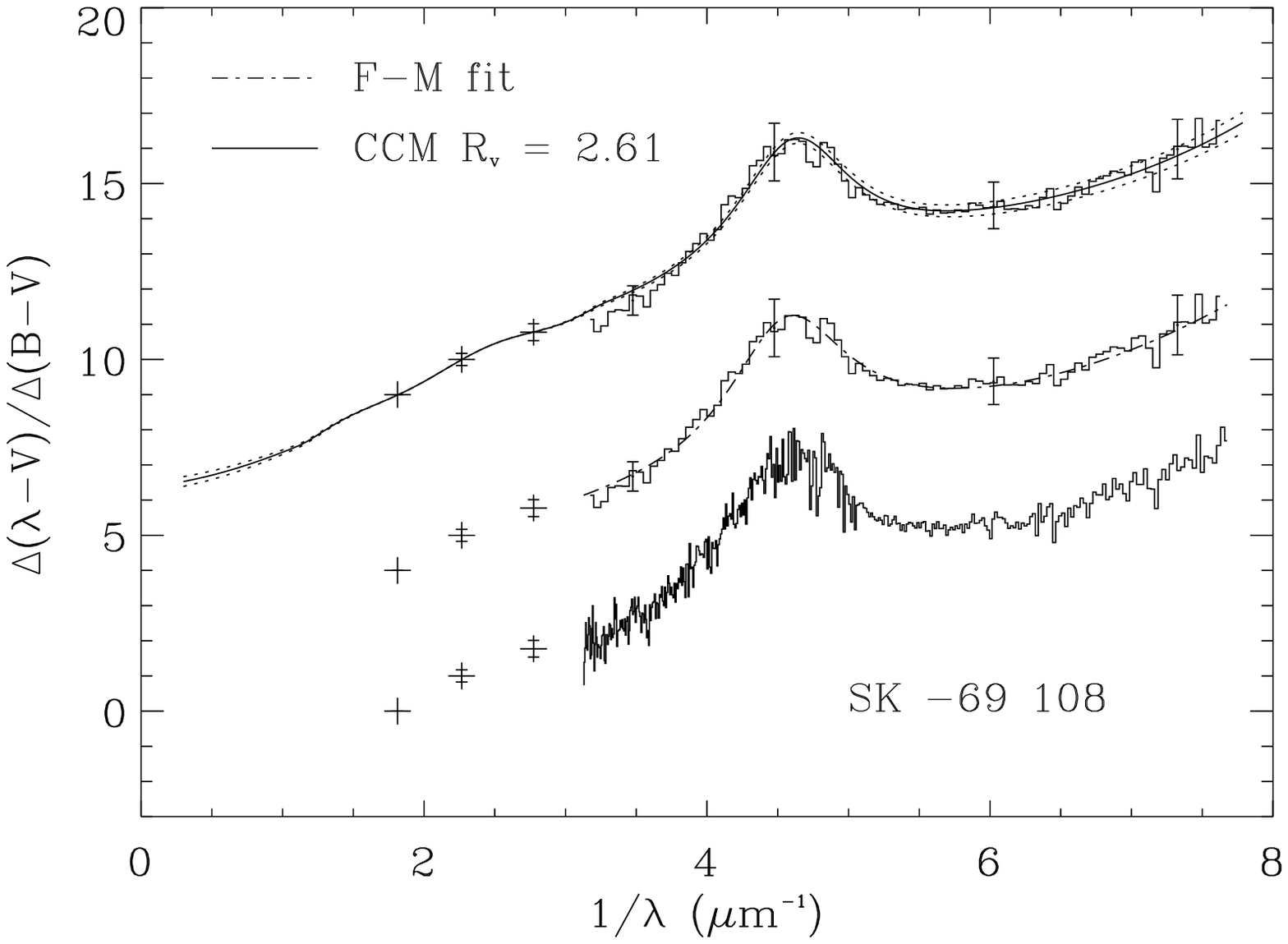}
                 {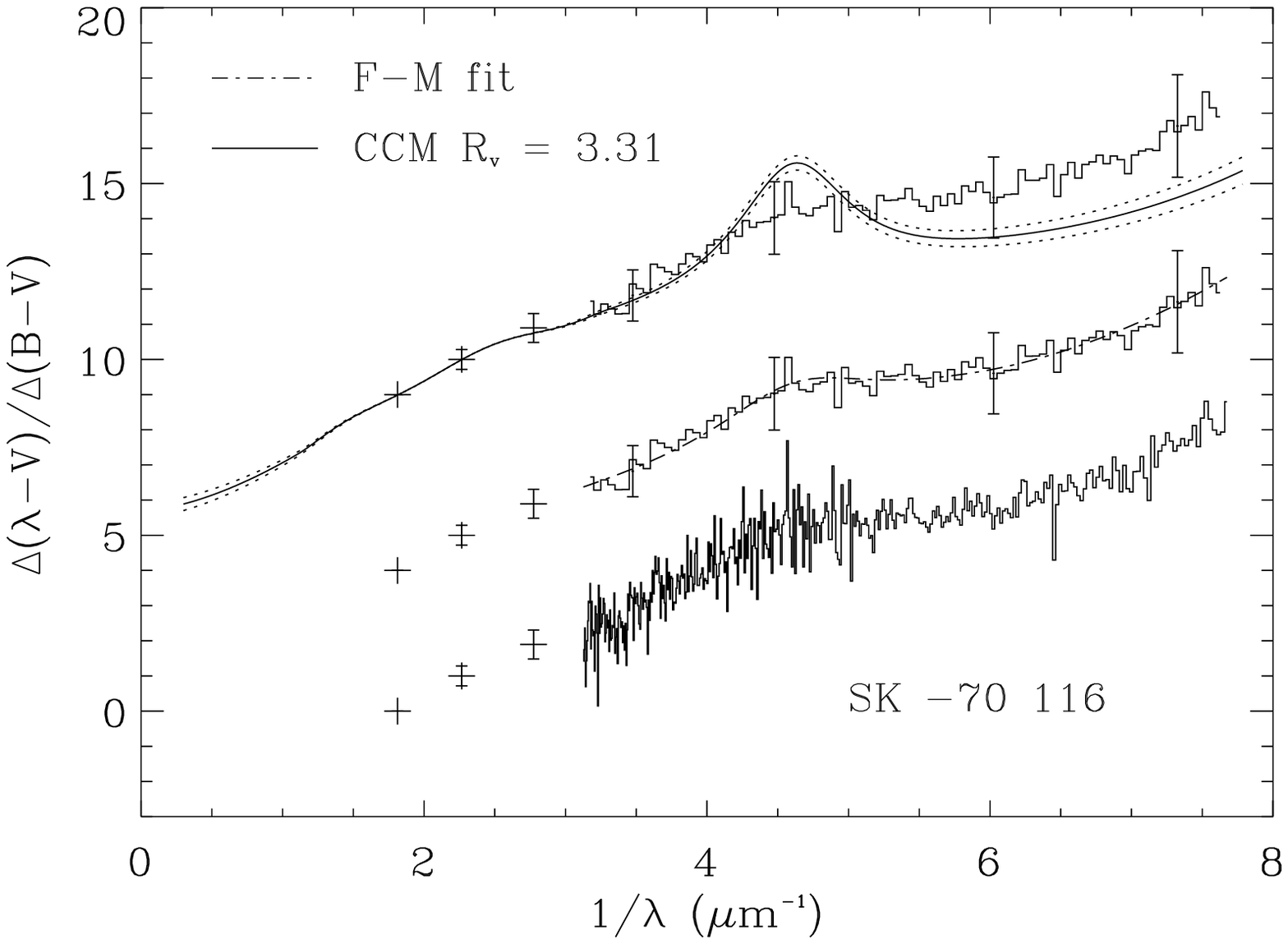}{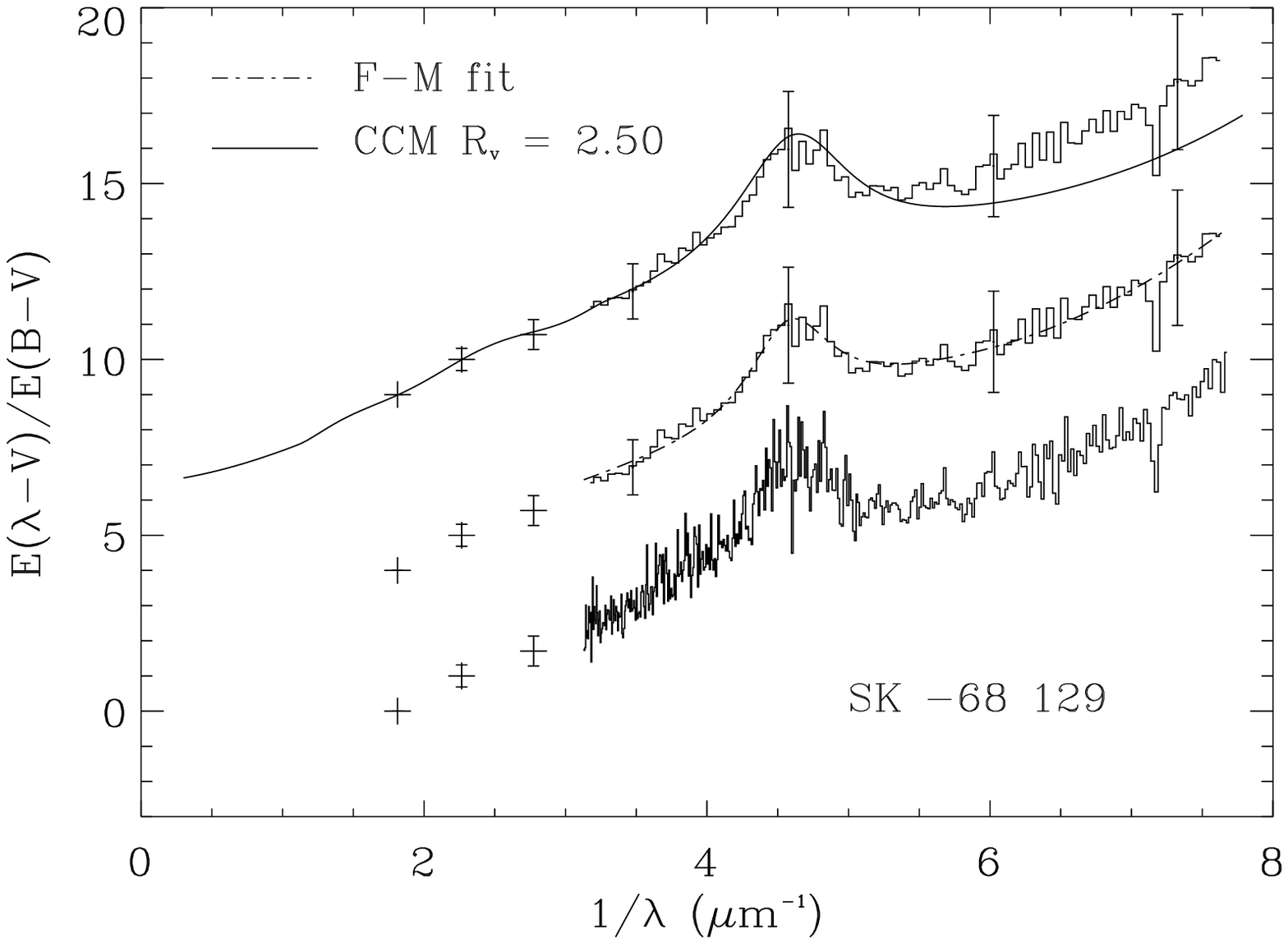}
\end{center}
\end{figure}
 
\begin{figure}[tbp]
\begin{center}
        \plotfour{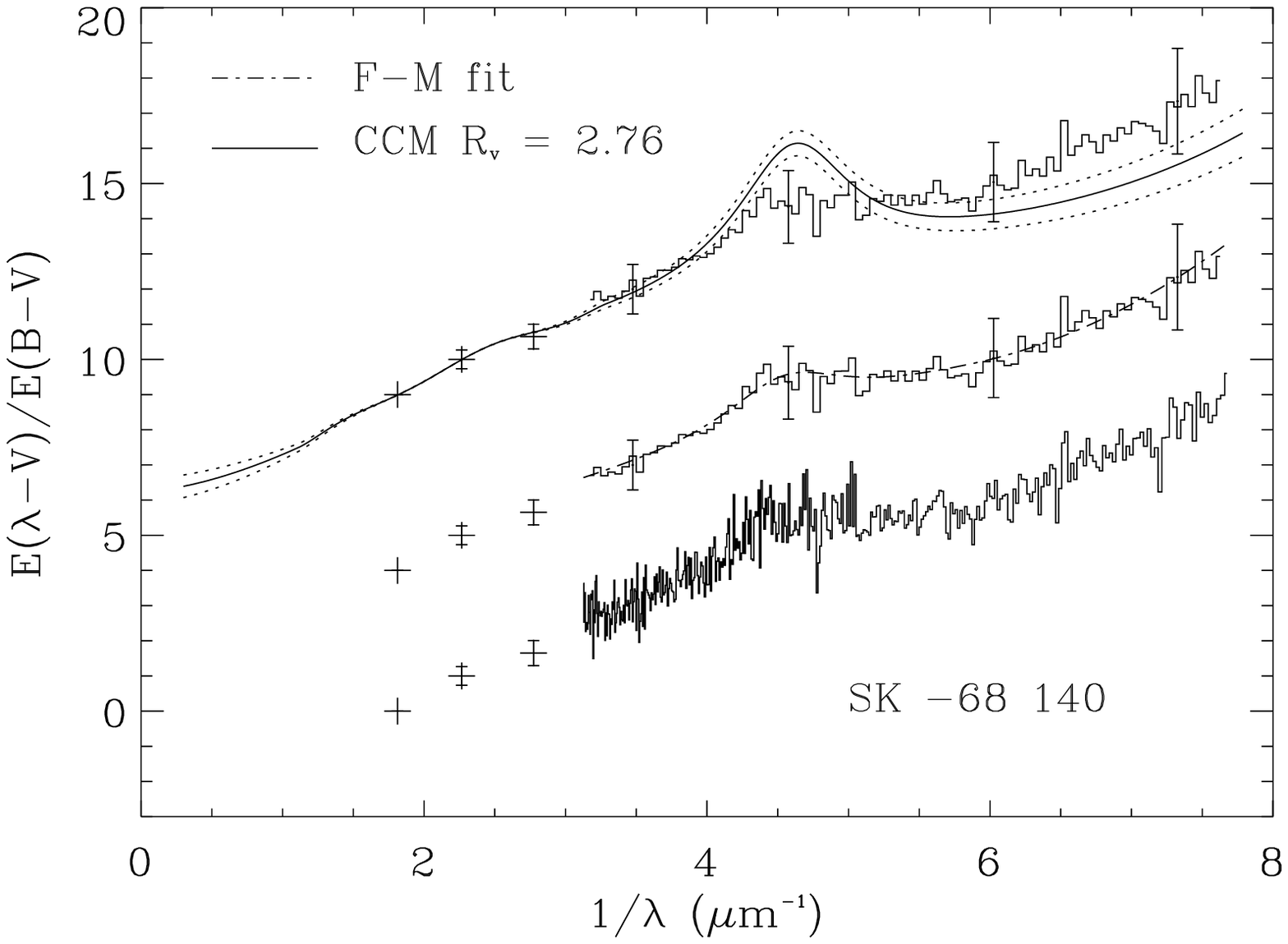}{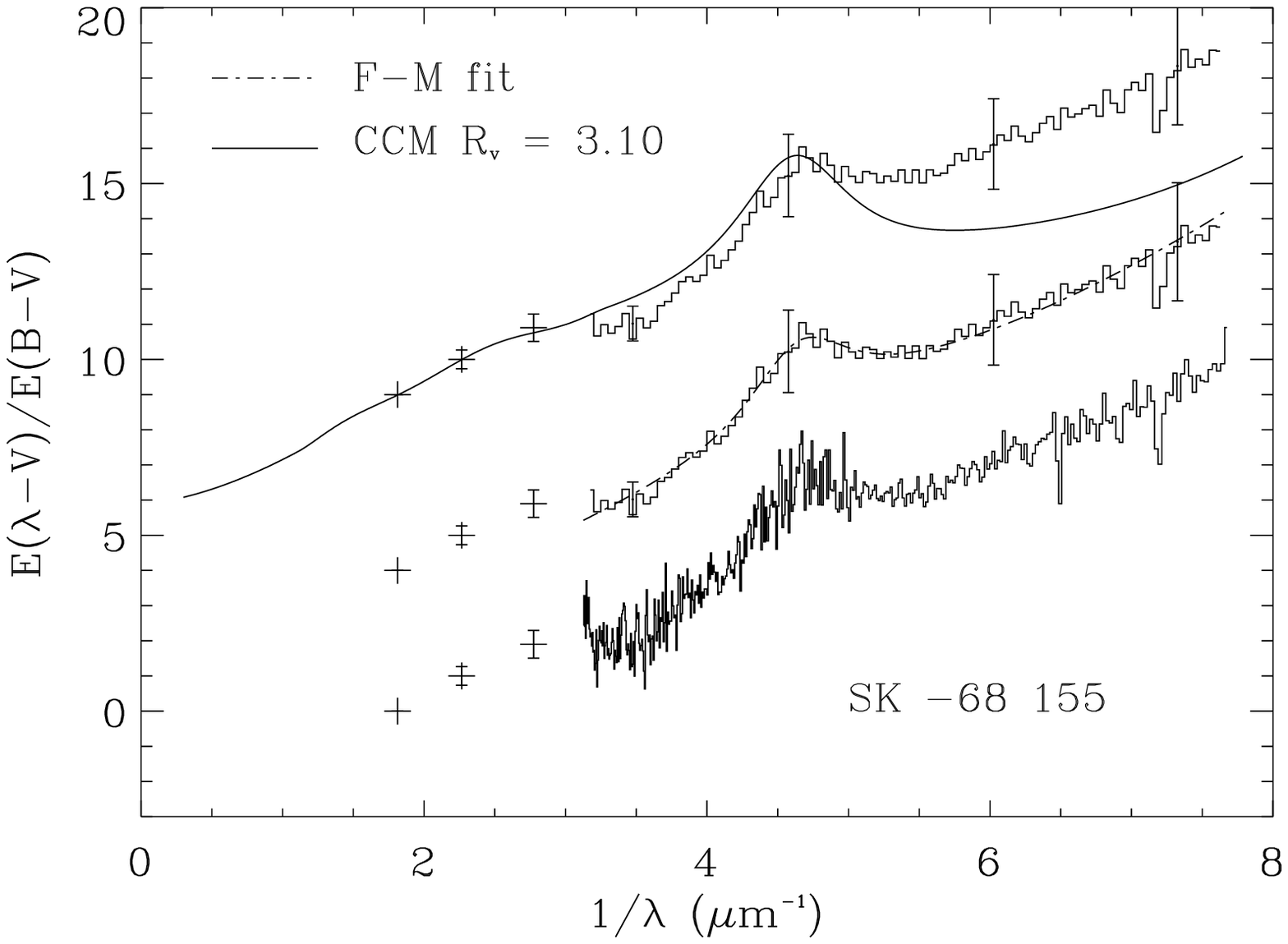}
                 {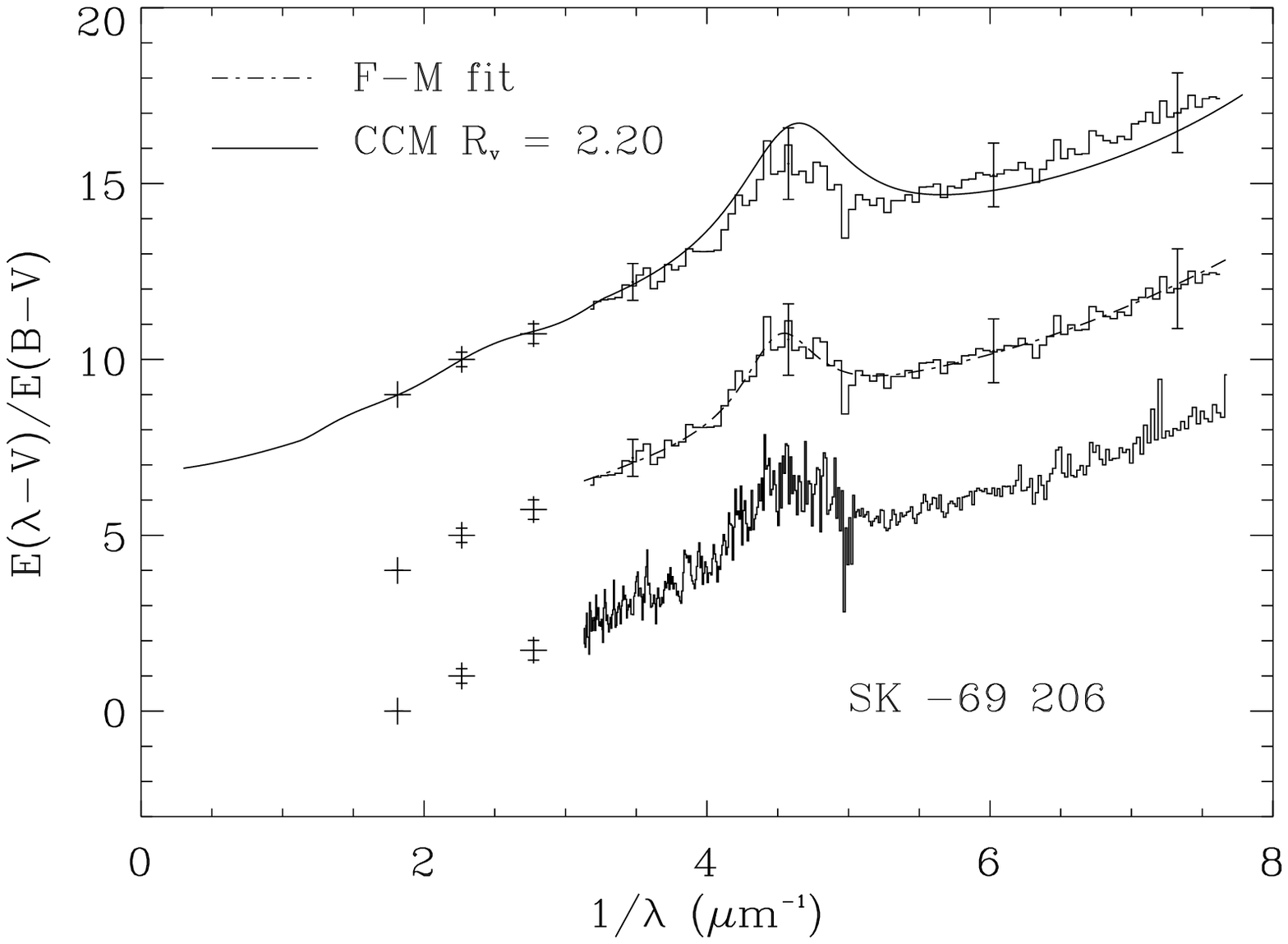}{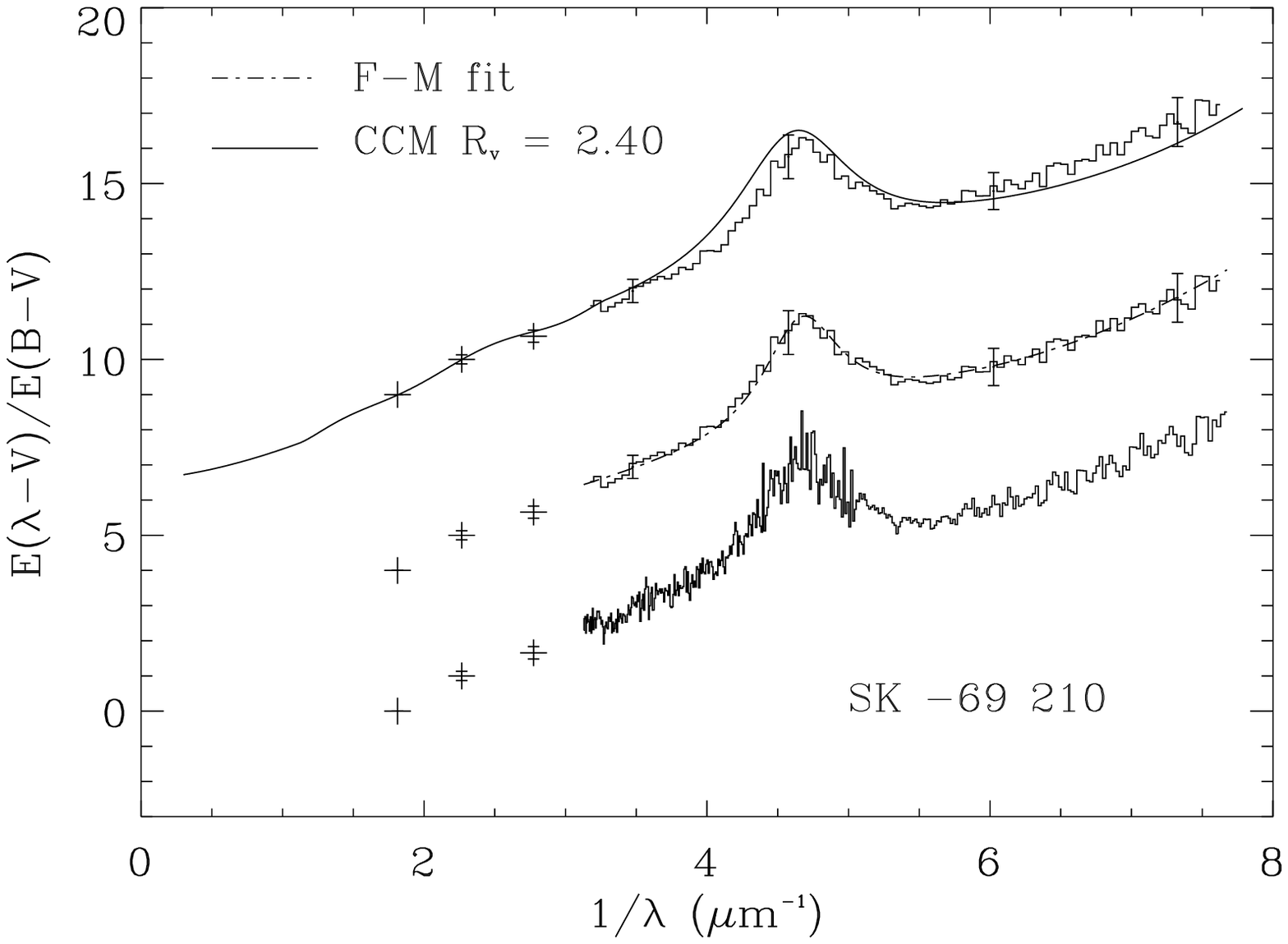}
                 {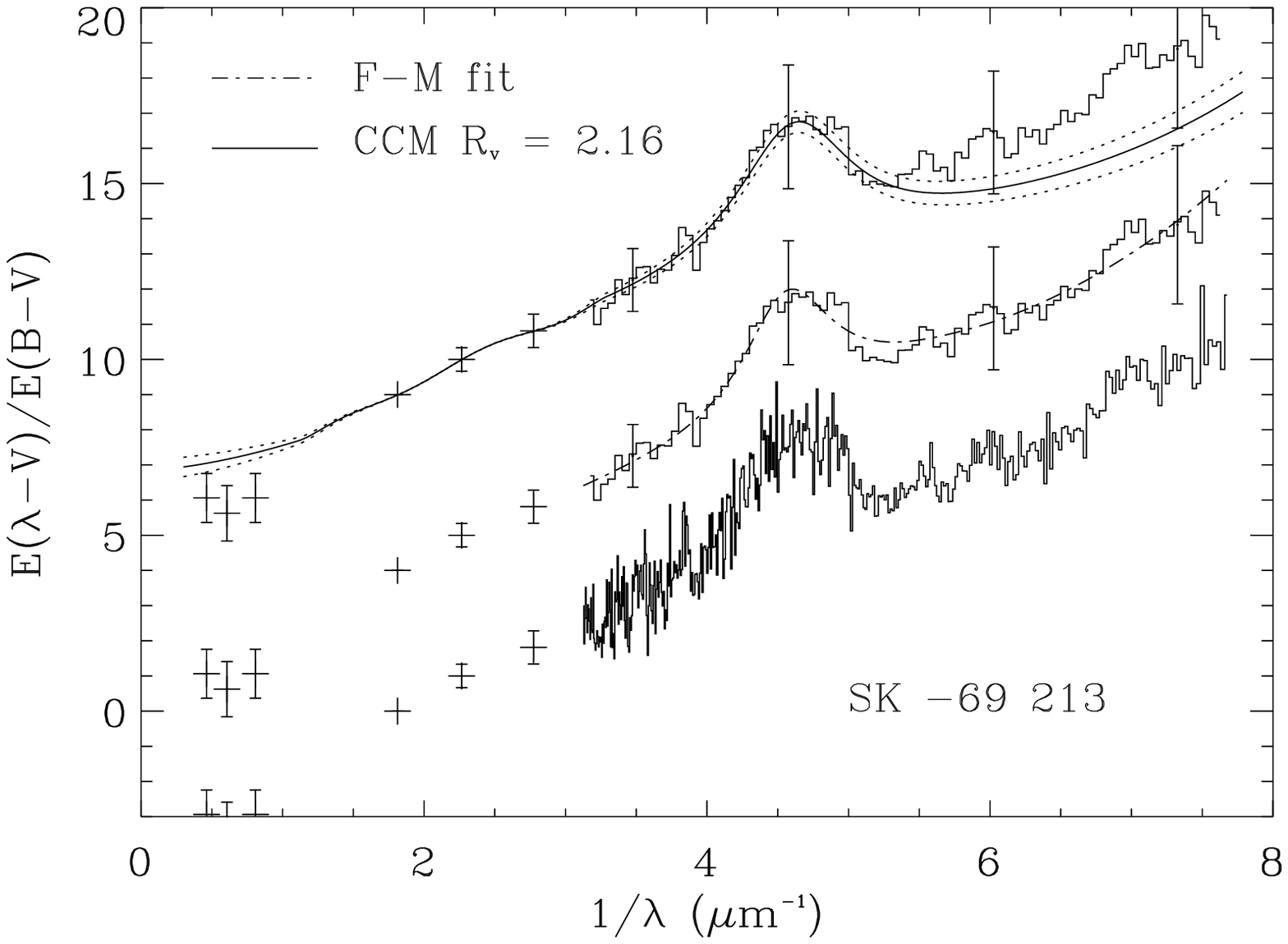}{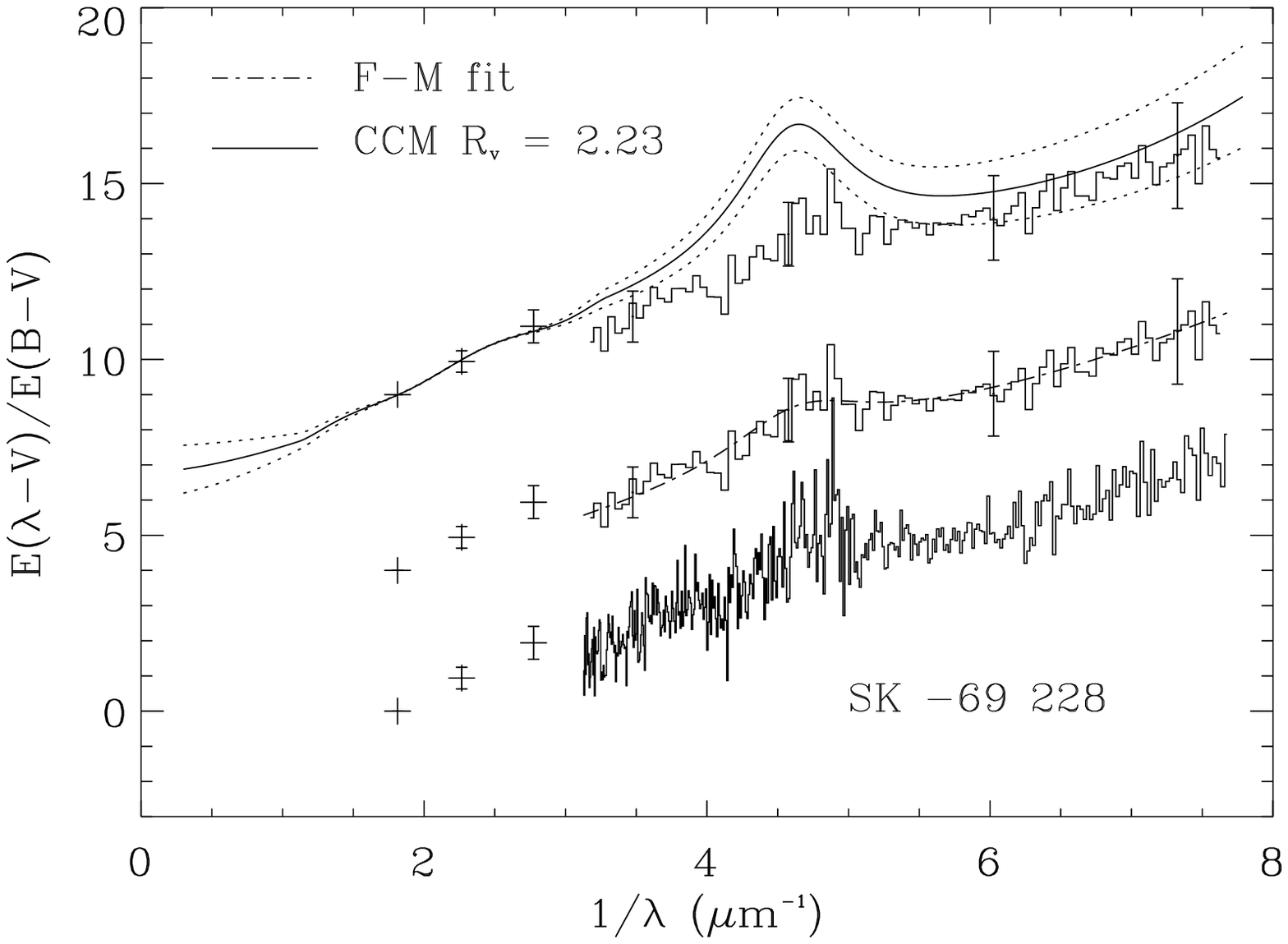}
                 {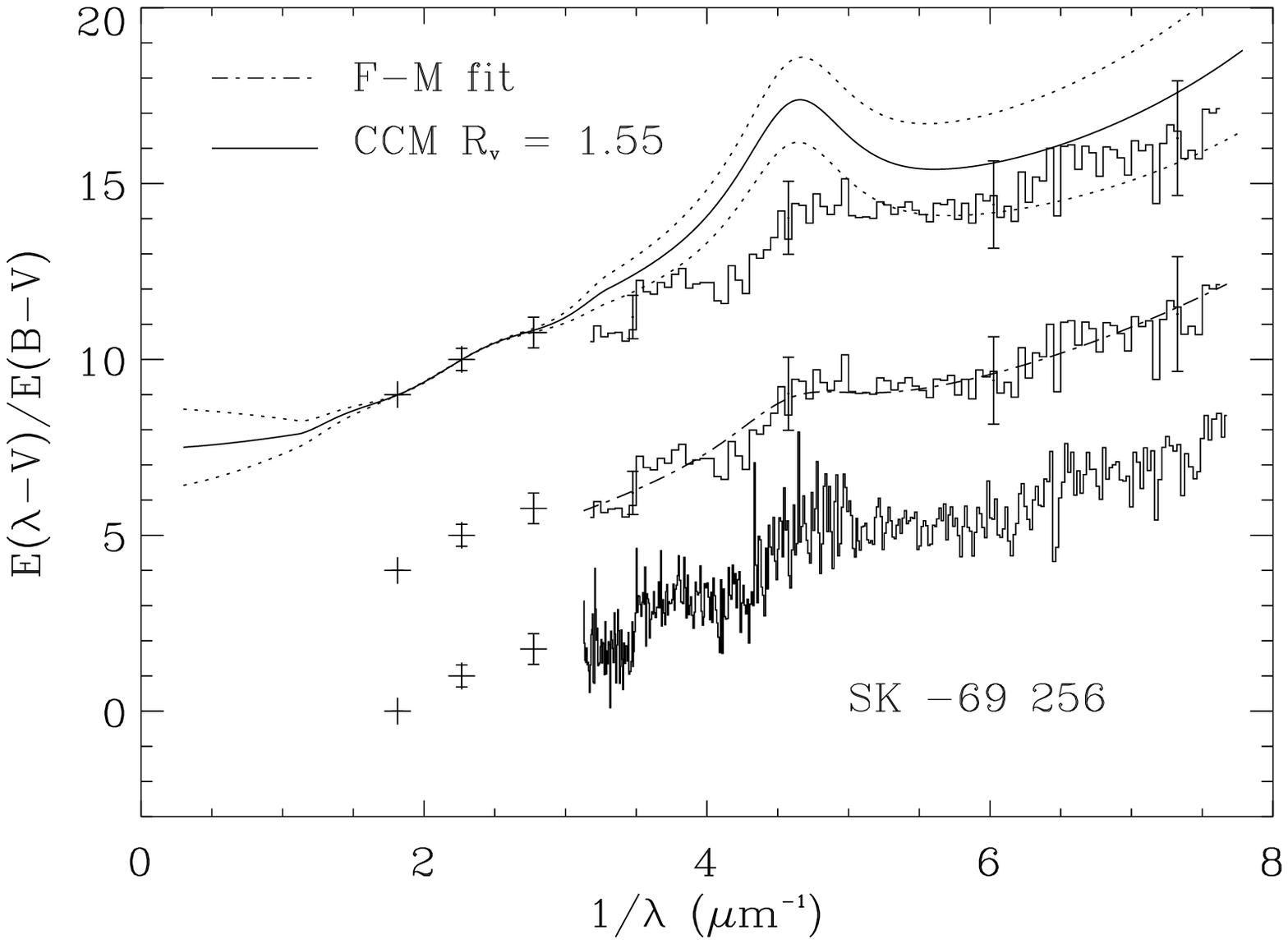}{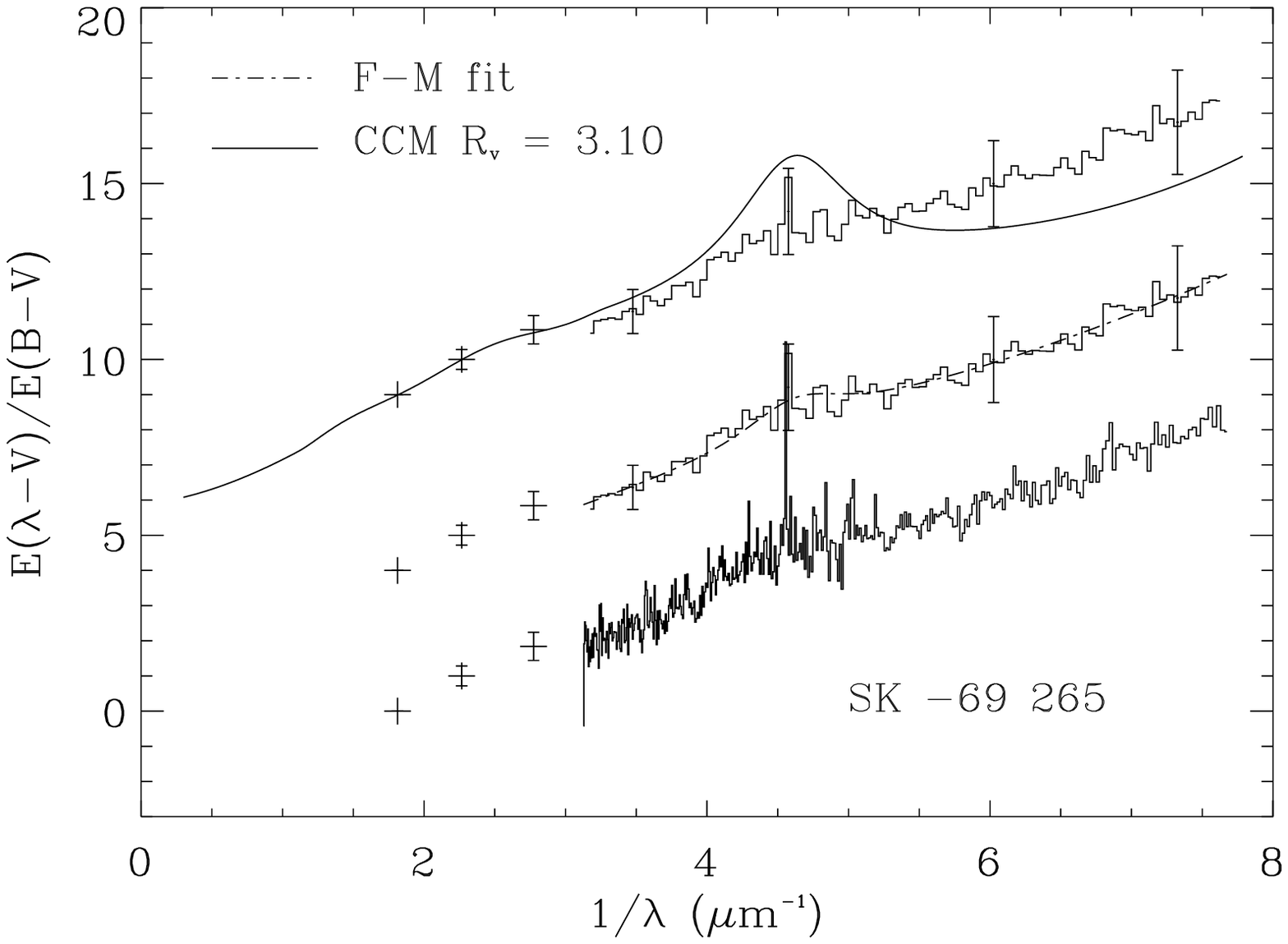}
\end{center}
\end{figure}
 
We determined $R_V$ values for all of the reddened stars in our sample which 
had R, I, J, H, and/or K observations.  Eleven reddened stars had measurements 
in at least three of these bands (Morgan \& Nandy 1982; Clayton \& Martin 1985).  
Intrinsic colors were taken from Johnson (1966) and Winkler (1997) assuming the 
reddened stars' UV spectral types.  The $R_V$ values were determined by assuming 
all extinction laws take the form of Rieke \& Lebofsky (1985) (CCM).  The 
uncertainties were calculated from the range of $R_V$ values which were 67\% 
probable using the reduced $\chi^2$ statistic (Taylor 1982).  The $R_V$ values 
and uncertainties are give in Table~4.
We do not include $-$69 256 in Table~4 or any of the subsequent analysis using
measured $R_V$ values as it value of $R_V$ is very uncertain (1.55$\pm$1.18).

\begin{deluxetable}{lcc}
\tablewidth{0pt}
\footnotesize
\tablecaption{Measured R$_V$ Values.}
\tablehead{
\colhead{SK} & \colhead{R$_V$} & \colhead{ $\sigma _{R_V}$ }
}
\startdata
$-$66 ~~19 & 2.46 & 0.25 \nl
$-$67 ~~~2 & 2.31 & 0.44 \nl
$-$69 108 & 2.61 & 0.15 \nl
$-$70 116 & 3.31 & 0.20 \nl
$-$68 140 & 2.76 & 0.35 \nl
$-$69 213 & 2.16 & 0.30 \nl
$-$69 228 & 2.23 & 0.74 \nl
%$-$69 256 & 1.55 & 1.18 \nl
$-$69 270 & 2.71 & 0.11 \nl
$-$69 279 & 2.43 & 0.31 \nl
$-$69 280 & 2.56 & 0.39 \nl
\enddata
\end{deluxetable}

\section{Discussion}

\subsection{Average Curves}
\subsubsection{30 Dor/Non--30 Dor}
\begin{figure}[tbp]
\begin{center}
\plotone{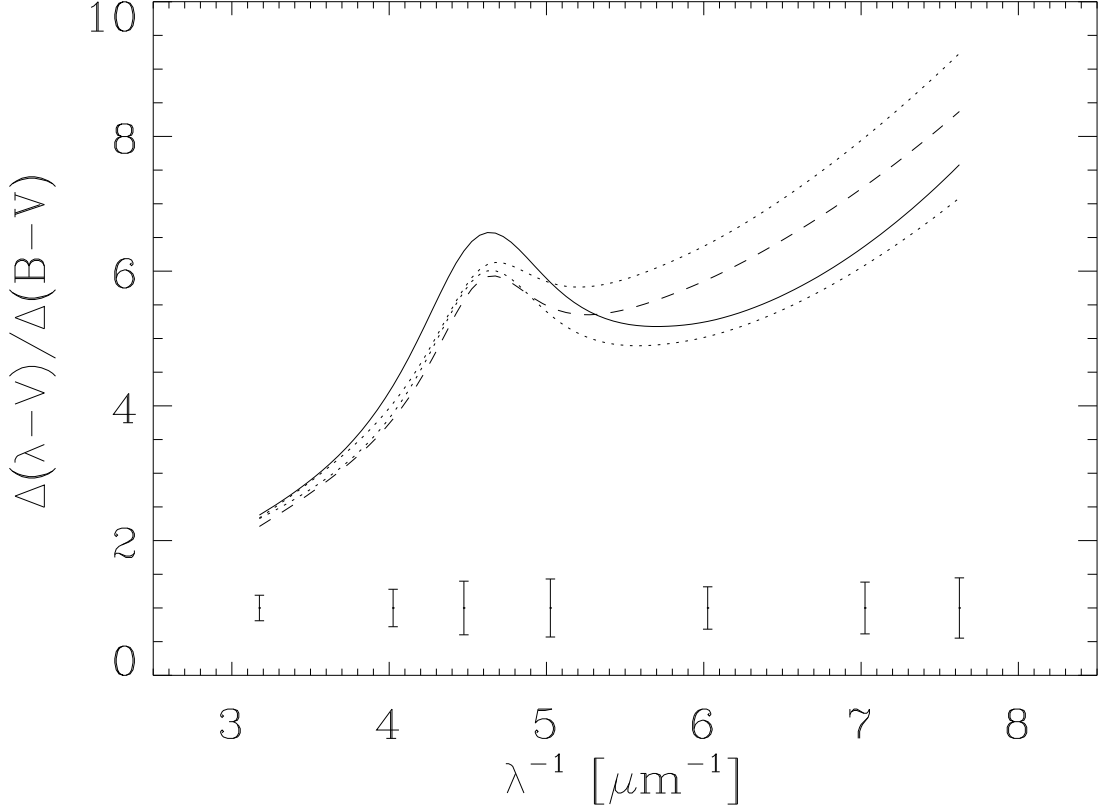}
\caption{Comparison of average 30 Dor (dashed line this study, upper dotted
line F86) and non--30 Dor (solid line this study, lower dotted line F86)
extinction curves. 1$\sigma$ error bars for the new 30 Dor and non--30 Dor average curves
have been plotted at various wavelengths. \label{fig_30dor_n30dor}}
\end{center}
\end{figure}

A very important result from previous work on the LMC was the apparent difference 
between UV extinction properties in the 30  Dor region and other sightlines in the 
LMC (Clayton \& Martin 1985; F85, F86). 
Reddened stars were assigned to the non--30 Dor
($d_{proj} \ge 1~kpc$, 7 objects) and 30 Dor
($d_{proj} < 1~kpc$, 12 objects)
samples based 
on their projected distance from R~136 as in previous studies.  
We have calculated average extinction curves
for our new 30 Dor and
non--30 Dor samples, weighting the individual curves by their uncertainties.
The FM parameters
of the average curves were calculated as the sample mean and the uncertainties
for the average FM parameters are the standard deviation of the mean 
for the respective samples, eg. $\sigma _{i}/\sqrt{N}$.
Formal FM fits to the average curves yielded identical parameters within the uncertainties.
In Figure~\ref{fig_30dor_n30dor}, the new average extinction curves for 30 Dor and non--30 Dor are 
shown with the results of F86 plotted for comparison. The extinction curves of 
F85 and F86 are virtually the same but their uncertainty estimates are quite different. 
At 7.0 $\micron^{-1}$, the difference between the Fitzpatrick 30 Dor and non--30 Dor 
curves is 1.86 $\pm$ 0.41 (F85).  In F86, the uncertainties are estimated to be about 
twice as large making the difference about 
2$\sigma$. Our results are similar to F86 but the 30  Dor curve is slightly lower and the 
non--30  Dor curve slightly higher in our averages.  We find the difference between 
the average curves at 7.0 $\micron^{-1}$ to be 0.89 $\pm$ 0.53. So the significance of 
differences in far--UV extinction between the 30 Dor and non--30 Dor samples is less,
being only slightly greater than 1.5$\sigma$.
The difference in bump strength between our 30 Dor and non--30 Dor 
samples is slightly more significant.
Our average non--30  Dor
bump strength ($A_{bump} = C_3/\gamma ^2$ = 2.97$\pm$0.30) is slightly larger
than that of F86 ($A_{bump} = 2.58$).
This is not unexpected as 
we have included two new lines of sight
with strong bumps in our non--30  Dor average.  In addition, the improvements
realized by using IUE spectra reduced with NEWSIPS are
most apparent near the bump.  
Our average 30 Dor bump strength ($A_{bump}=2.12\pm0.20$) is only 
slightly larger
than that found by F86 ($A_{bump}=1.86$).
The difference in bump strength between our 30 Dor and non--30 Dor 
samples is $\Delta A_{bump}=0.85 \pm 0.36$, slightly greater than $2\sigma$.

\subsubsection{LMC~2/LMC--general}
\begin{figure}
\begin{center}
\plotone{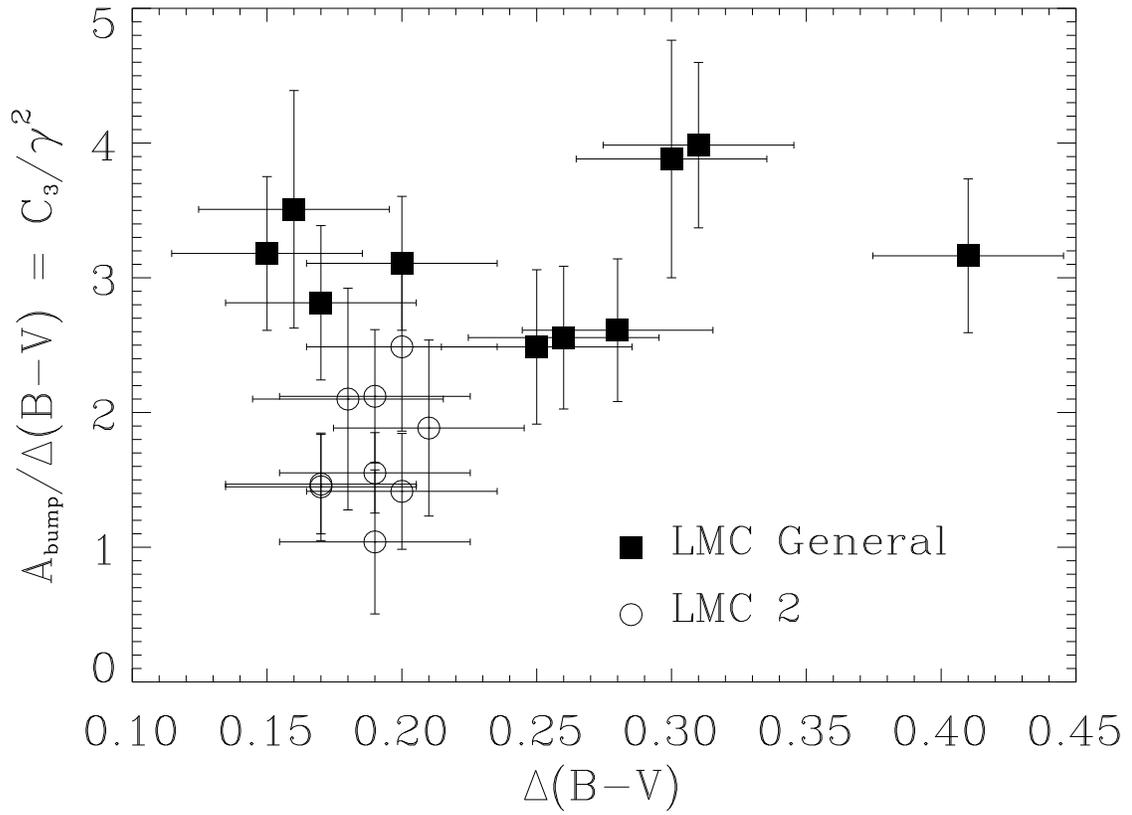}
\caption{Plot of the bump strength normalized to $\Delta$(B$-$V) vs. $\Delta$
(B$-$V).  Symbols
represent the samples discussed in the text. \label{fig_bs_ebv}}
\end{center}
\end{figure}

However, the conclusion drawn by F86 that there are significant intrinsic variations 
between extinction curves within each of the 30  Dor and non--30  Dor 
samples is strengthened by the additional lines of sight included in this study. 
In the non--30 Dor sample, for instance, SK $-$68 23 has a strong bump and SK $-$70 116 
has almost no bump. Similar differences are seen in the 30 Dor sample. 
To try and isolate a sample of sightlines with weak bumps, we have plotted bump strength 
versus $\Delta$(B$-$V) in Figure~\ref{fig_bs_ebv}.  We discovered that there is a group of stars
with similar reddenings (0.17 $\leq \Delta(B-V) \leq$ 0.21) and bump strengths that also 
lie close together in the LMC.  These 
stars lie in or near the region occupied by the supergiant shell LMC 2 on the southeast 
side of 30 Dor (Meaburn 1980; see Figure~\ref{fig_ha_map}). 
This structure, which is 475 pc in radius,  
was formed by the combined stellar winds and supernovae explosions from the stellar 
association within (Caulet \& Newell 1996).
There are nine stars in the LMC 2 group, eight of which are from the 30  Dor sample 
and one (SK $-$70 116) from the non--30  Dor sample. 
Four 30 Dor stars (SK $-$68 129, $-$69 206, $-$69 210 and $-$69 213) are removed from
our new LMC~2 sample.
These four stars lie in or near 
a prominent dust lane separating
the 30  Dor star formation region from the LH~89 and NGC~2042 stellar associations;
SK $-$69 206 is on the south--eastern edge of the dust lane near the 
stellar association LH~90 while SK $-$69 210 is in the middle of the
dust lane, coincident with CO clouds 7 \& 8 of Johansson et. al. (1998).
While located in the traditional 30  Dor region, these sightlines 
have strong bumps, typical of the non--30 Dor dust.

\begin{figure}[tbp]
\begin{center}
\plotone{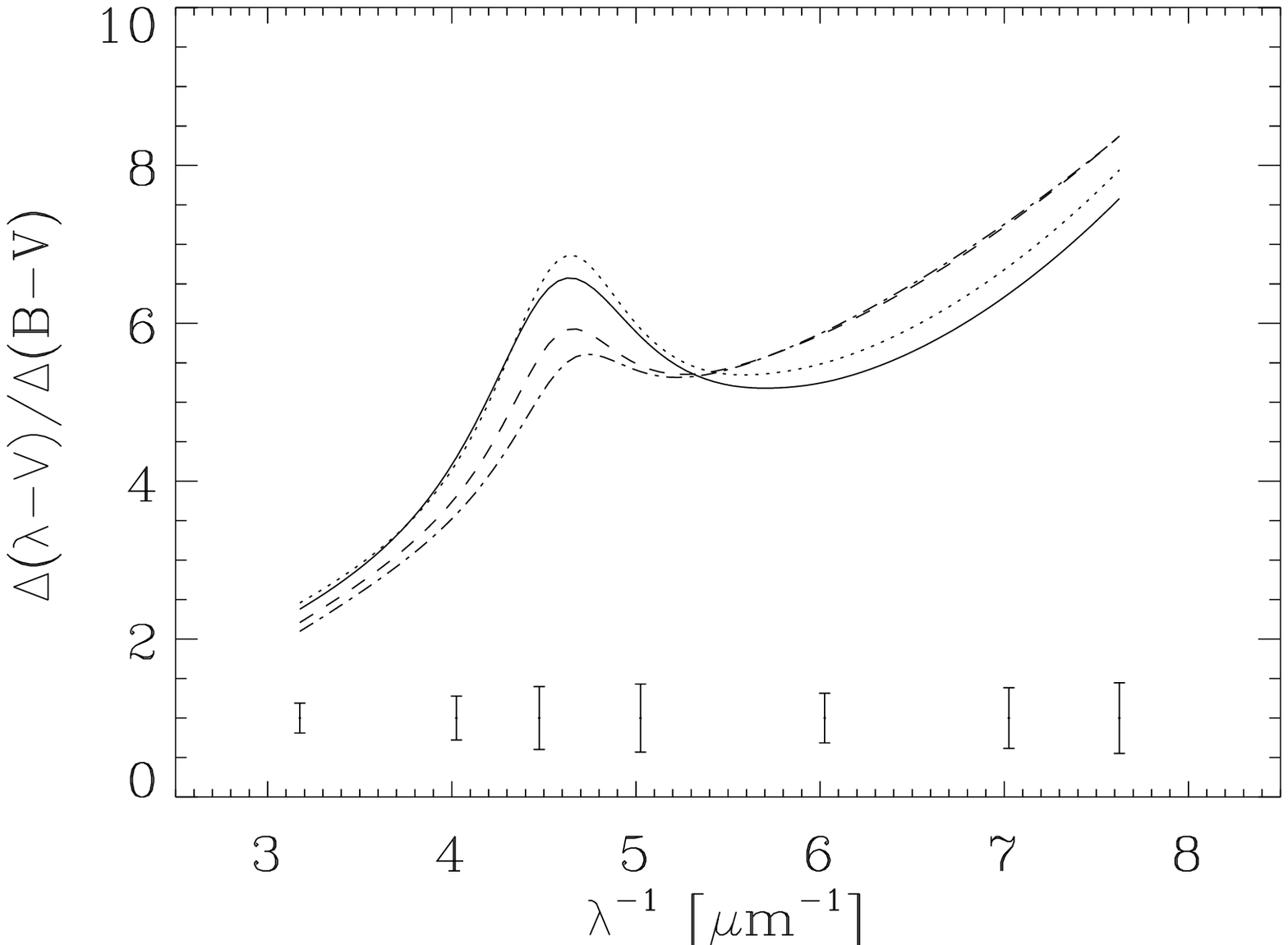}
\caption{Comparison of 30 Dor/non--30 Dor average curves from this study
(dashed line and solid line, respectively) with the LMC average and LMC 2
average curves discussed in the text (dotted line and dot-dash line,
respectively). 1$\sigma$ error bars for the new 30 Dor and non--30 Dor average curves
have been plotted at various wavelengths. \label{fig_lmc2_lmcave}}
\end{center}
\end{figure}

\begin{figure}[tbp]
\begin{center}
\plottwo{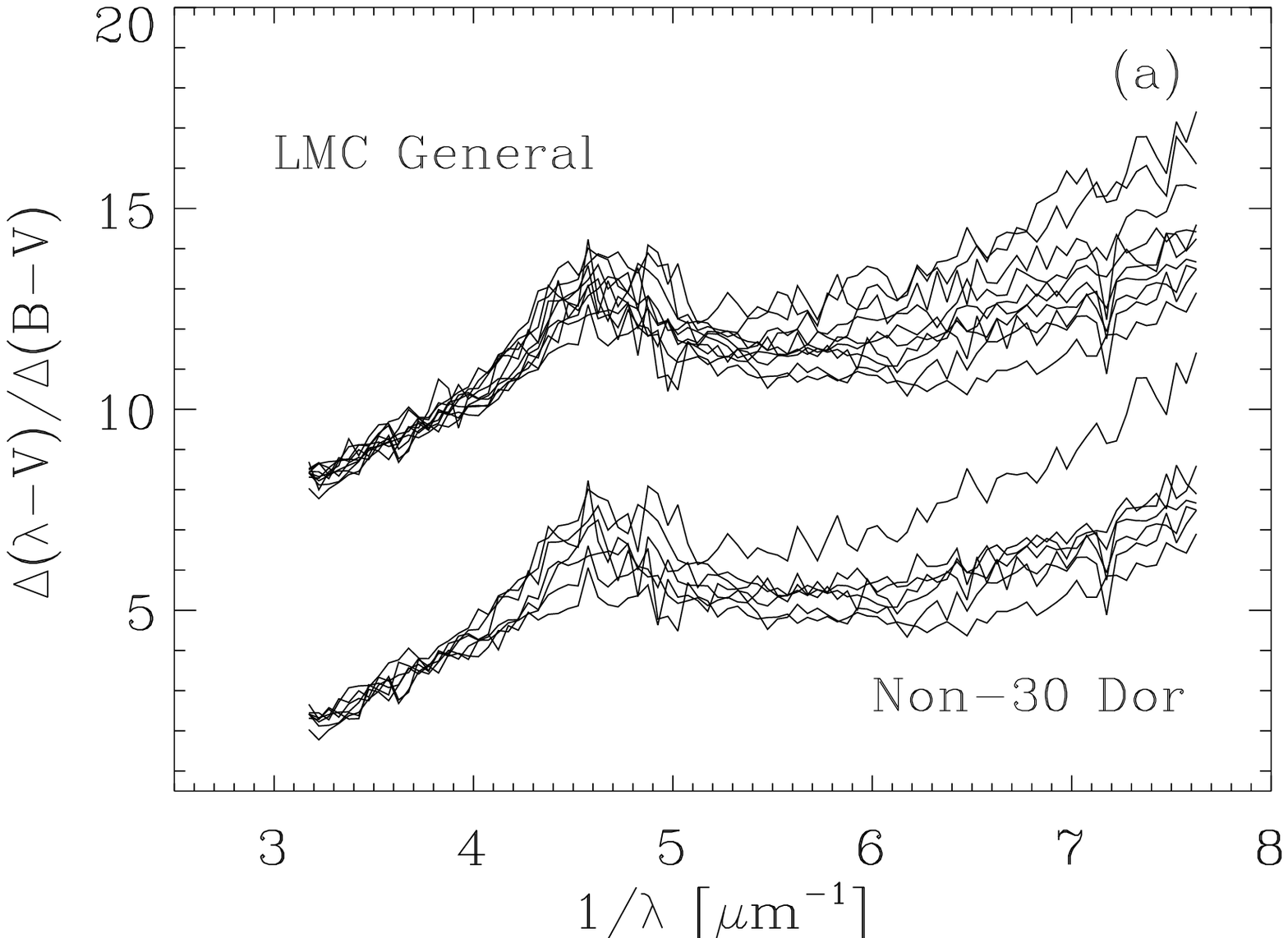}{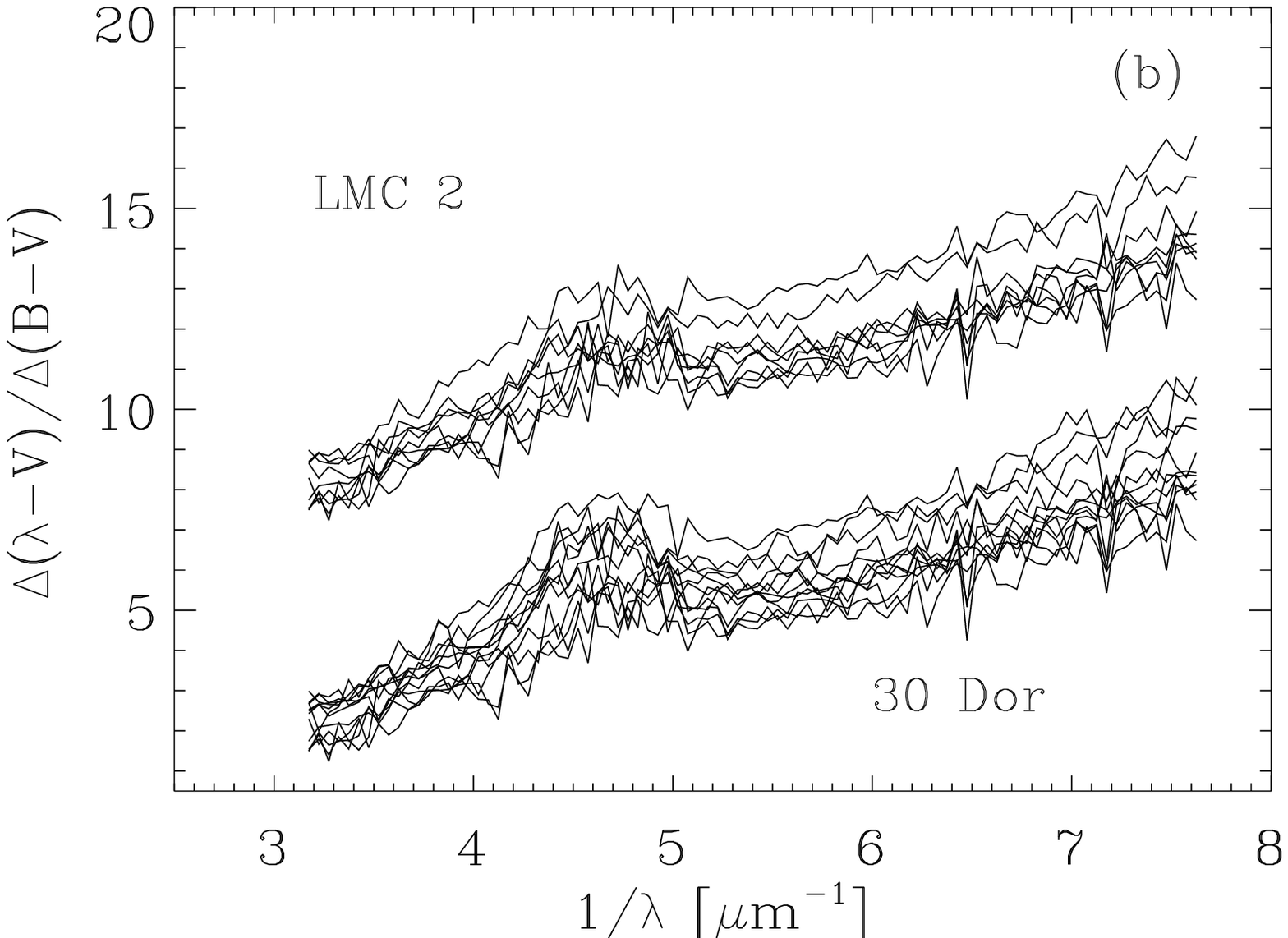}
\caption{(a) Comparison of the individual curves within the non--30  Dor sample
(lower curves) and LMC--general sample (upper curves).  The LMC--general curves
have been offset 6 units for clarity. (b) Same as (a) for the 30  Dor sample
(lower curves) and LMC 2 sample (upper curves).  The LMC 2 curves have been offset
by 6 units for clarity. \label{fig_dispersion}}
\end{center}
\end{figure}

An average extinction curve has been calculated for the LMC~2 stars and also for the
remaining ten stars which we will call LMC--general.  FM parameters and their respective
uncertainties were calculated as above and are reported in Table 3.  The parameters for the
average Galactic curve
as derived from FM are also shown for comparison.
The average curves for LMC 2 and LMC--general samples 
are plotted in Figure~\ref{fig_lmc2_lmcave}.  
These two curves 
show a very significant difference in bump strength ($\Delta A_{bump}$ = 1.41 $\pm$ 0.21) 
but the far--UV curves lie within one sigma of each other.  It is worth noting that the
average Galactic bump strength is very similar to that of the LMC--general sample.
In Figure~\ref{fig_dispersion}
we have over--plotted individual curves within each sample.
The dispersion about the mean 
bump strength is significantly less for both the LMC--general sample compared to
the non--30  Dor sample (0.50 and 0.78, respectively; Figure~\ref{fig_dispersion}a) 
and for the LMC 2 sample 
compared to the
30 Dor sample (0.43 and 0.72 respectively; Figure~\ref{fig_dispersion}b).  

\subsection{Variations Within the Samples}
\begin{figure}[tbp]
\begin{center}
\plottwo{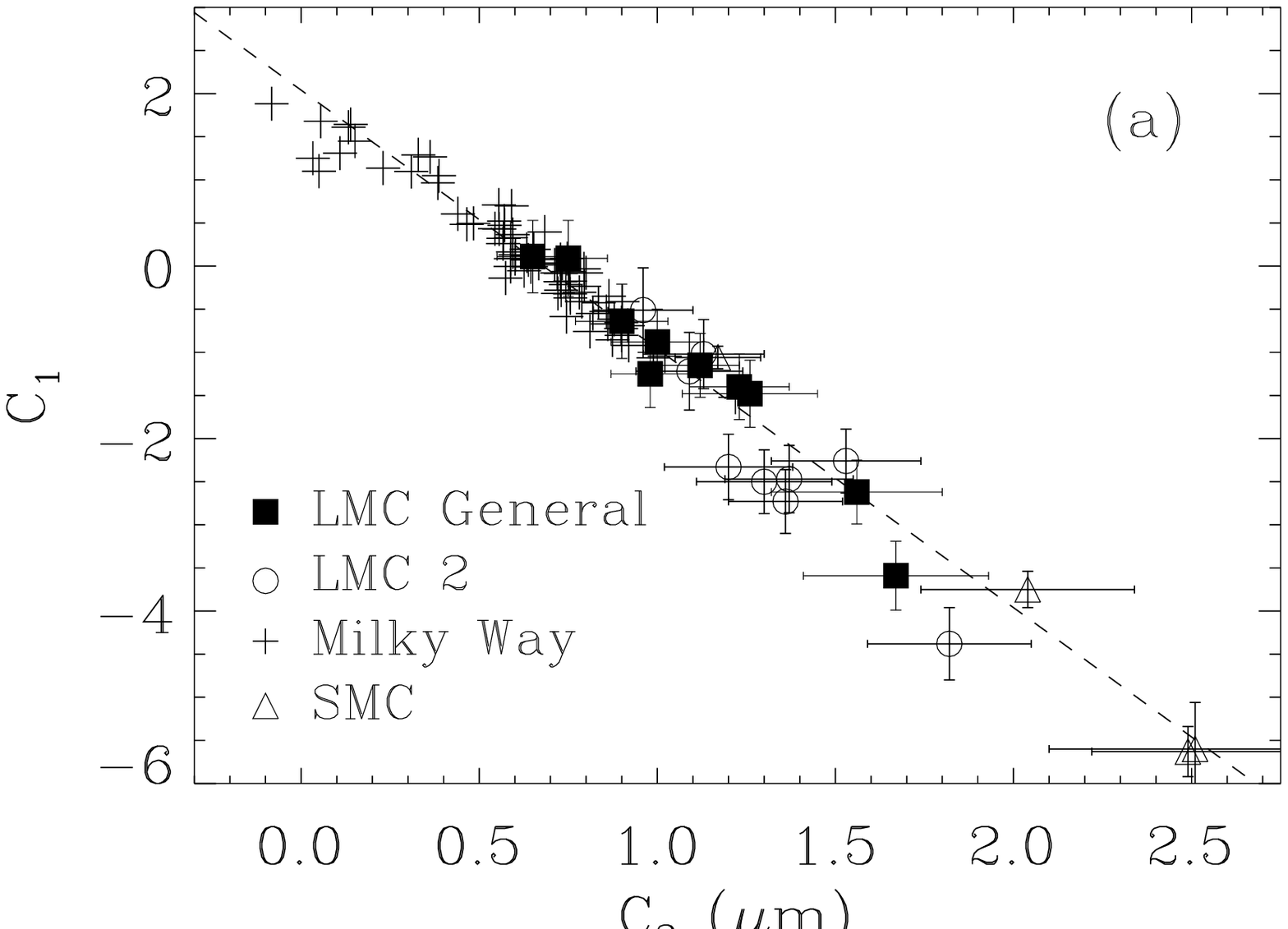}{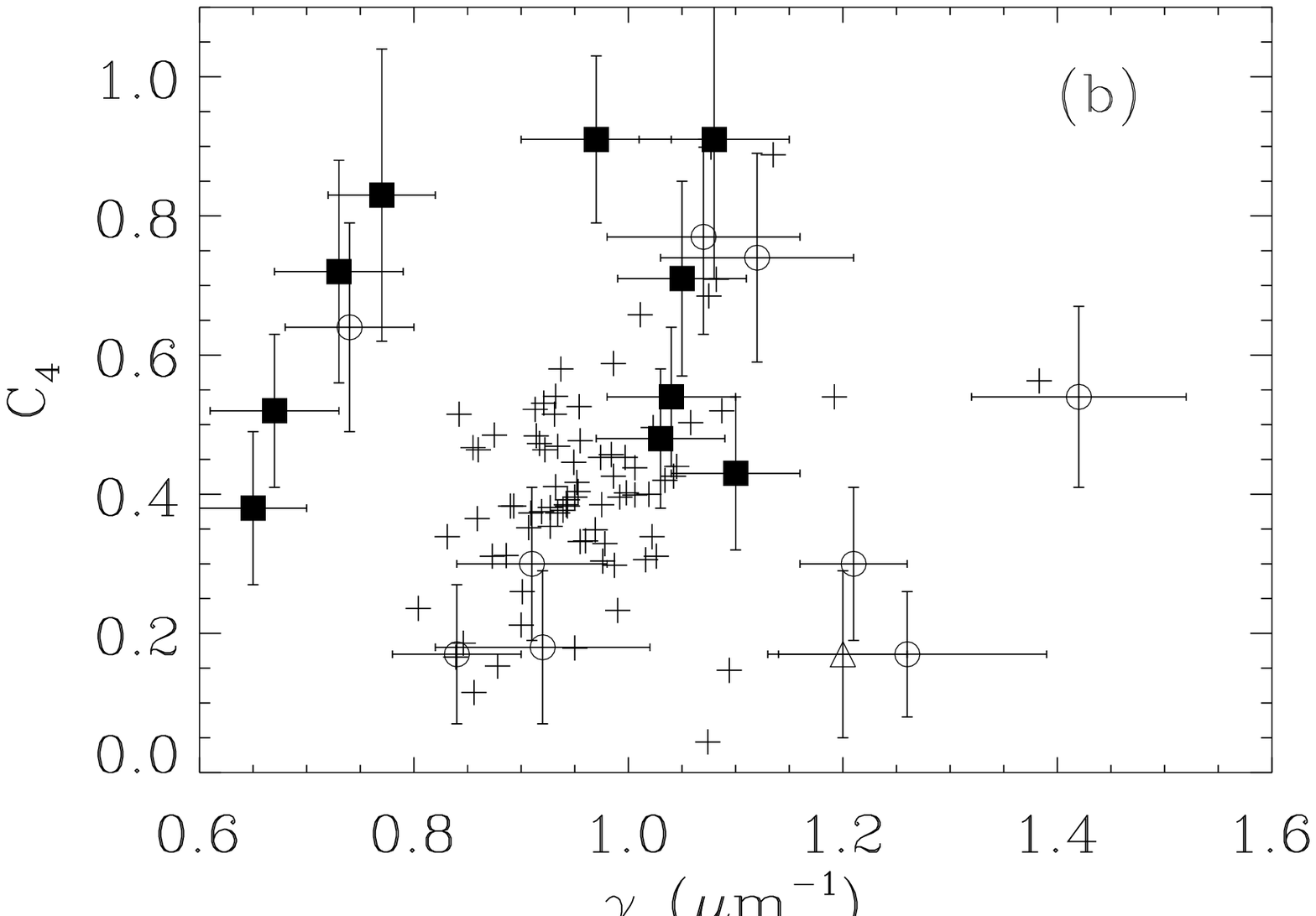}
\caption{(a) Plot of $C_1$ vs. $C_2$ for the LMC--general sample, 
LMC~2 sample, the SMC,and the Galaxy.  The dashed line is the least--squares fit the Galactic data
given by Fitzpatrick \& Massa (1988).  (b) Plot of $C_4$ vs. $\gamma$. Symbols are as in
Figure 7a.  In both figures, Galactic data from FM, SMC data from GC. \label{fig_FMparam}}
\end{center}
\end{figure}

The form of the UV extinction, as parameterized by FM, 
along a given line of sight is potentially
a powerful diagnostic of the dust grains responsible for the extinction but
the physical interpretation of variations and correlations among the FM
parameters is unclear.
However, to the degree that they represent underlying physical processes it
is useful to examine them within our two LMC samples.
The coefficients of the linear component of the UV extinction ($C_1 + C_2x$) are
not independent in the Galaxy (Fitzpatrick \& Massa 1988) but are in fact themselves
linearly related.  Fitzpatrick \& Massa (1988) interpret the relationship as either
a single grain population modified by evolutionary processes or a varying mixture of several
grain populations with different UV extinction slopes or a combination of both factors.
While $C_1$ and $C_2$ have similar values between the two LMC samples, both
LMC samples exhibit systematically smaller values of $C_1$ and systematically larger
values of $C_2$ relative to the Galaxy.  In the SMC, the values of $C_1$ and $C_2$
are even more extreme than in the LMC (GC).
However, the values of $C_1$
and $C_2$ for all these galaxies follow the same 
linear relationship (see Figure~\ref{fig_FMparam}a).
Hence, whatever underlying physical processes
or dust components
are responsible for the variations in the linear part of the UV extinction must
operate similarly in the Galaxy, the LMC and the SMC.
Fitzpatrick \& Massa (1988) suggested a possible correlation between the
FM parameters $C_4$, which measures the far UV curvature, and $\gamma$, the bump
width. In Figure~\ref{fig_FMparam}b we plot $C_4$ against $\gamma$ for the Galaxy,
the LMC, and the SMC. Only one SMC sightline (AzV~456) is included since 
the remaining three sightlines have no bump and $\gamma$ is undefined (GC).
There is no correlation between these parameters in the LMC extinction data.
The physical significance of $C_4$ is unclear; the far UV extinction
is a combination of the linear term and the $C_4$ polynomial term and the separation
is mathematical rather than physical (CCM).
However, such a correlation may arise if $C_4$ and $\gamma$ are due to different 
grain populations provided that the different populations
respond to environmental factors in a similar way (Fitzpatrick \& Massa 1988).
This is consistent with the conclusion
of CCM that the processes producing changes in extinction must be efficient over a range
of particle sizes and compositions.  In this case, the absence of correlation in the LMC
would suggest that environmental processes are affecting the different grain populations
differently.  

The FM parameters $C_1$, $C_2$, $C_3$, and $C_4$
depend on $R_V$ and so interpreting
relations among them in the absence of $R_V$ information is difficult.
CCM found that the general shapes of the UV extinction curves in the Galaxy,
expressed as $A_{\lambda}/A_{V}$,
are well represented by a one parameter family of curves characterized by the 
value of $R_V$.
It is of interest to determine whether the UV extinction in the LMC follows the
relation of CCM and whether the deviations from CCM in the LMC can be 
related to deviations seen in the Galaxy
We will discuss the FM bump parameters ($x_0$ and $\gamma$) separately in \S 3.2.2.

\subsubsection{CCM and the LMC}
\begin{figure}
\begin{center}
\plotone{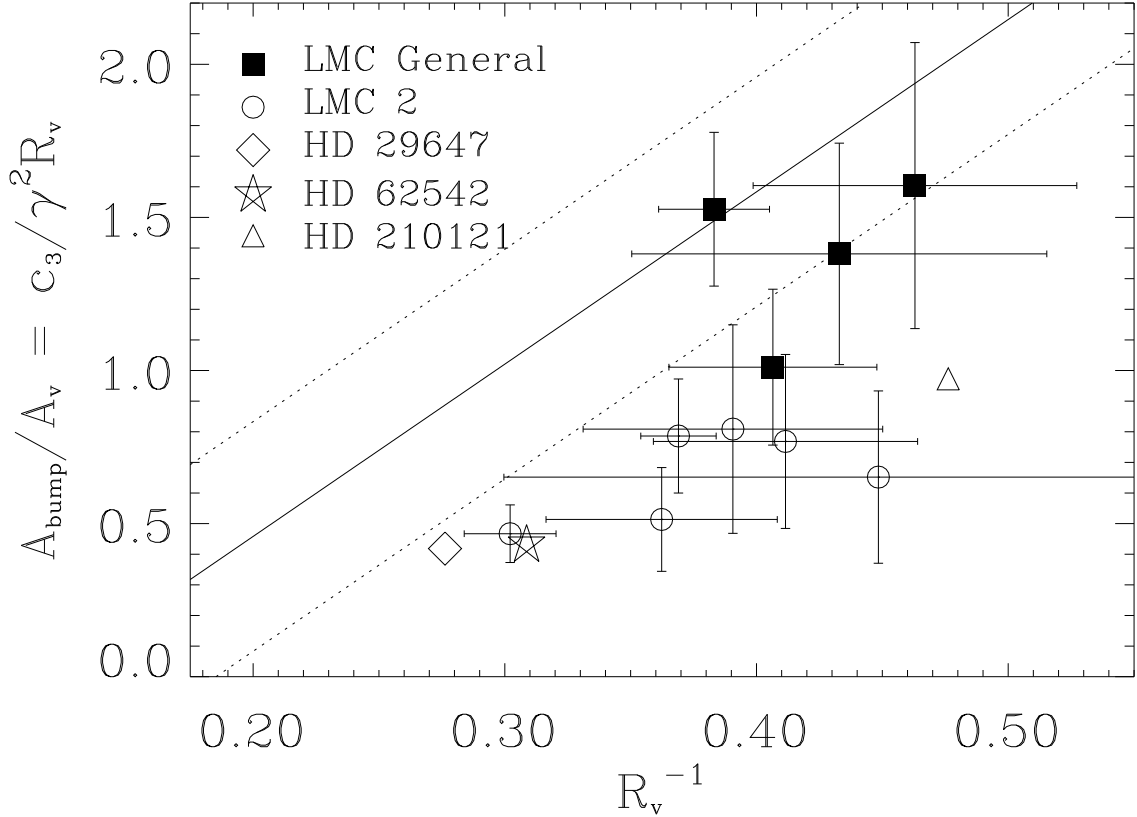}
\caption{Bump strength normalized to A$_V$ plotted vs. R$_V^{-1}$ for LMC stars with measured
values of R$_V$.  The solid line represents the mean CCM relationship and the dotted lines the
approximate dispersion around the mean for the CCM sample of Galactic stars.
Symbols represent the samples
discussed in the text. For comparison, several
Galactic stars with ``unusual'' extinction curves are plotted. Bump strengths for the Galactic
stars were taken from Cardelli \& Savage (1988) (HD 29647) and Welty \& Fowler (1992) (HD 62542
\& HD 210121).  R$_V$ values are from Messinger et. al. (1997) (HD 29647),
Whittet et. al. (1993) (HD 62542) and
Larson et. al. (1996) (HD 210121). \label{fig_bs_rv}}
\end{center}
\end{figure}

\begin{figure}
\begin{center}
\plotone{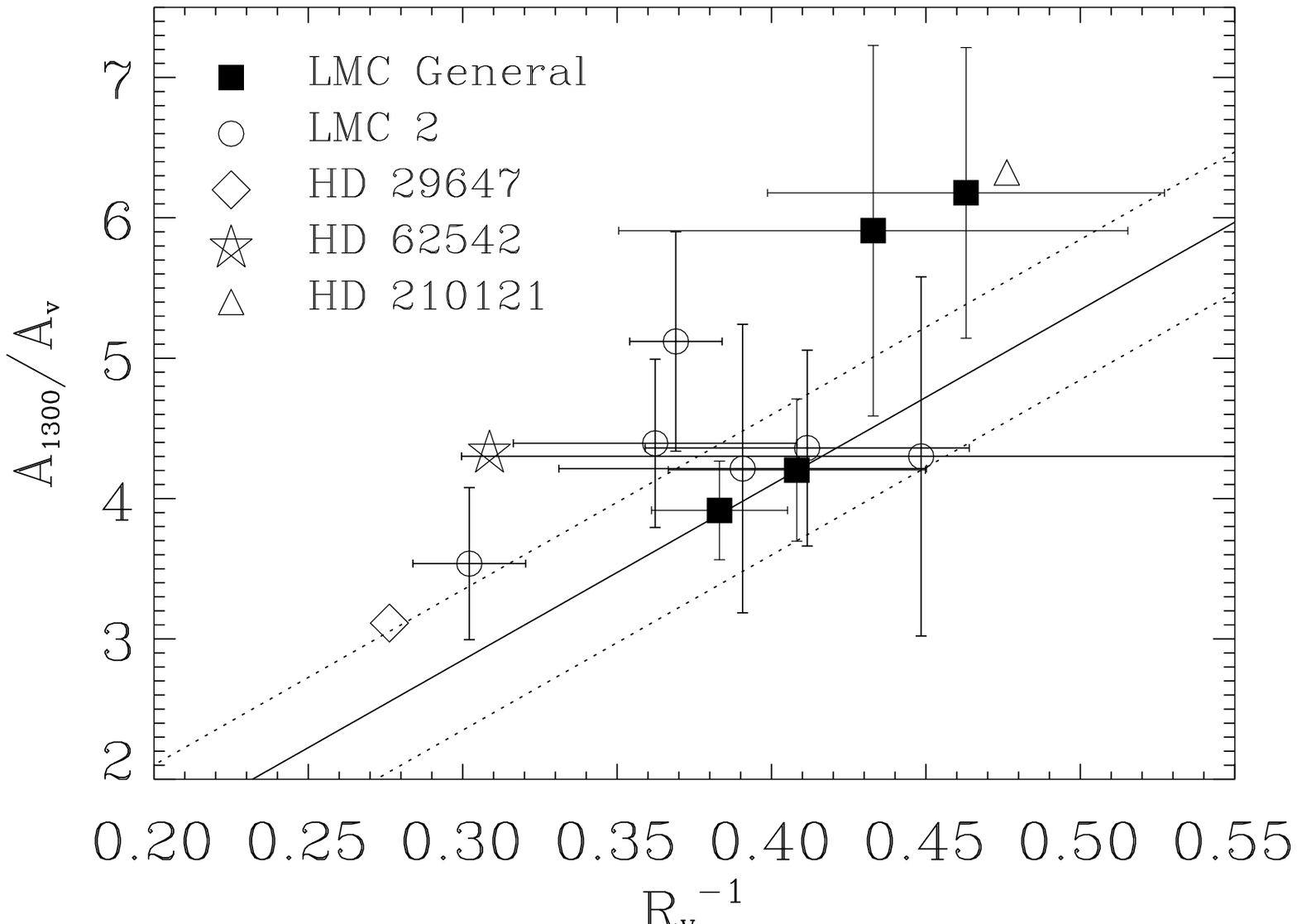}
\caption{Plot of the extinction ratio $A_{1300}/A_V$ vs. $R_V^{-1}$ where
$A_{1300}$ is the extinction at $\lambda=1300$~\AA\ plotted as in Figure~\ref{fig_bs_rv}.
\label{fig_a1300_rv}}
\end{center}
\end{figure}

There is an average Galactic extinction relation, $A_{\lambda}/A_{V}$, over
the wavelength range 0.125 $\micron$ to 3.5 $\micron$, which is applicable
to a wide range of interstellar dust environments, including lines of
sight through diffuse dust, dark cloud dust, as well as that
associated with star formation (CCM;
Cardelli \& Clayton 1991; Mathis \& Cardelli 1992).  The
existence of this relation, valid over a large wavelength interval,
suggests that the environmental processes which modify the grains are
efficient and affect all grains. The CCM
relation depends on only one parameter
$R_V$, which is a
crude measure of the size distribution of interstellar dust.
Only eleven LMC sightlines in our sample have measured values of $R_V$. Seven of these
are in the LMC 2 sample.  The CCM curves for these eleven stars are plotted in 
Figure~\ref{fig_ext_curves}.
The LMC 2 curves cannot be fit by a CCM curve with
any value of $R_V$ because of their weak bumps.
The average LMC--general curve is very similar to a Galactic CCM extinction curve with $R_V=2.4$.
However, only four stars in this sample have measured $R_V$ values so it is not clear how well
their extinction curves follow CCM.  
SK $-$67 2 and $-$69 213 have stronger FUV extinctions than their respective CCM curves
while SK $-$66 19 appears too weak in the bump.  Only SK $-$69 108 clearly follows the
CCM relationship.
In Figures~\ref{fig_bs_rv} and \ref{fig_a1300_rv} we plot bump strength and $A_{1300}/A_V$ versus
$R_V^{-1}$ for ten stars with measured $R_V$'s; SK $-$69 256 is excluded due to its very
uncertain $R_V$ value. Figure~\ref{fig_bs_rv} shows that bump
strength is consistent with CCM for the LMC--general sample while the LMC 2 sample
has bumps which fall below the typical CCM values.  
Very little can be said about the relationship with $R_V$ in the far UV
as seen in Figure~\ref{fig_a1300_rv}.  
The uncertainties are quite large and though both the LMC 2 and LMC--general sample
appear to be consistent with CCM in the UV, they are also consistent with no $R_V$
dependence.  More accurate values of $R_V$ along more sightlines must be obtained before
it can be determined whether a CCM--like relationship may hold in the LMC.
According to CCM, $C_3$ and therefore $A_{bump}$ are proportional
to $R_V$ (Mathis \& Cardelli 1992).
However, since $C_3$(LMC--general)/$C_3$(LMC 2) = 2.25, that would imply that the average value
of $R_V$ should be more than twice as large in the LMC--general sample if a CCM--like
relationship exists.  There is
no indication from the available data that this is true.  In fact, the sightlines in both
samples appear to have low values of $R_V$ relative to the Galaxy (Table 4).  This may indicate that
dust grains in the LMC may be systematically smaller than in the Galaxy.

Although the general shape of the UV extinction in the Galaxy is well represented
by the $R_V$ parameterization of CCM, significant deviations are seen,
both in the far UV and the bump (Cardelli \& Clayton 1991; Mathis \& Cardelli 1992;
Fitzpatrick 1998).
There are well known Galactic sightlines which deviate from CCM in much the
same way that the LMC 2 sample does. Three deviant Galactic stars are plotted
in Figures~\ref{fig_bs_rv} and \ref{fig_a1300_rv} for comparison. 
HD 29647, 62542, and 210121 all show
weak bumps and strong far--UV extinction for their measured values of R$_V$ 
(3.62, 3.24 \& 2.1, respectively; Messinger et. al 1997; Whittet et. al. 1993; Larson et. al. 1996).
The bumps seen for HD 29647 and 62542 are not just weak but they
are very broad and shifted to the blue (Cardelli \& Savage 1988).  The unusual
extinction curve characteristics along these lines of sight have been attributed 
to their dust environments which are quite diverse.  
The dust toward HD 62542 has been swept up by bubbles blown by 
two nearby O stars and has been subject to shocks while
the HD 29647 sightline passes through a very dense, quiescent 
environment (Cardelli \& Savage 1988).  HD 210121 lies behind a single
cloud in the halo. There is no present activity near this cloud although
it was ejected into the halo at some time in the past (Welty \& Fowler 1992; 
Larson et. al. 1996).
These deviations from CCM in the Galaxy indicate that something
other than the size distribution of dust grains as measured by $R_V$
must be important in determining extinction properties along
a given line of sight.  
Evidently, the same is true in the LMC.  
Even though all the LMC sightlines have similar, low values of $R_V$, they
exhibit a variety of extinction curves.

\subsubsection{The Bump}
While the physical significance of the linear and far UV functions in the FM
parameterization is unclear,
the Drude profile fitting function for the 2175 \AA\ absorption bump 
which is part of the FM parameterization
does have some physical
significance as the expression of the absorption cross section of a damped
harmonic oscillator (CCM; FM; Mathis \& Cardelli 1992).  Further, neither 
$x_0$ or $\gamma$ depend on $R_V$ and so variations in these parameters are
directly tied to variations in the grains responsible for the bump feature.

There is no evidence for a systematic shift in the central position of the
bump in either LMC sample.  The weakness of the bump in the LMC~2 sample means
that $x_0$ and $\gamma$ are not strongly constrained in that sample.  In the 
LMC--general sample, there are no systematic redward shifts of the bump but three
sightlines are significantly shifted to the blue. 
The range of variation in the LMC is consistent with
that seen in the Galactic sample.  Several Galactic lines of sight, eg.
HD 62542 and HD 29647 have bumps that are significantly shifted to shorter
wavelengths ($x_0 = 4.74$ and 4.70, respectively; Cardelli \& Savage 1992).  
Several possibilities have been suggested to account for this including 
mantling of the grains and hydrogenation (Cardelli \& Savage 1992; Mathis 1994).

\begin{figure}
\begin{center}
\plotone{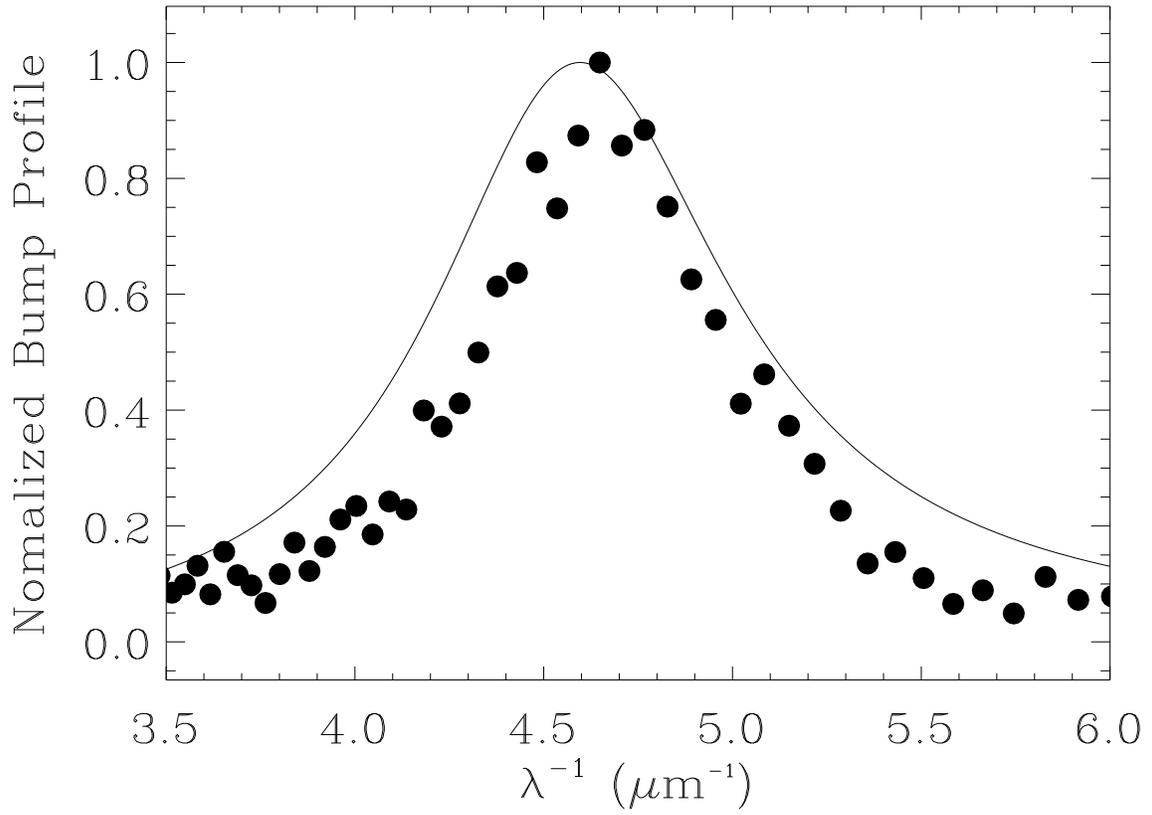}
\caption{Drude profile of SK $-$69 210 (filled circles) binned to $\sim$30\AA\ resolution
compared to the average Galactic Drude profile. \label{fig_drude210}}
\end{center}
\end{figure}

As in the Galaxy, there is real variation in the width of the bump between
various lines of sight in the LMC (Table 3).
Five lines of sight in our sample have bump widths which nominally
fall below the narrowest Galactic bump ($\gamma=0.8$, FM).  Of these five sightlines
two (SK $-$69 213 and SK $-$69 280) are affected by spectral
mismatches in the bump region. 
The true bump widths for these two sightlines are not
well constrained by the FM fitting procedure.  The remaining three narrow bump sightlines
(SK $-$68 129, SK $-$69 206, SK $-$69 210),
all in the LMC--general sample, interestingly all fall in or near the dust lane
on the northwest edge of 30 Dor.  The bumps
are well defined and the narrowness of the bump is real.
An expanded view of the SK $-$69 210 profile can be seen Figure~\ref{fig_drude210}.
There is a strong relationship between environment and
$\gamma$ in the Galaxy.  The narrowest bumps are associated with bright nebulosity
while wide bumps are associated with dark, dense clouds (Cardelli \& Clayton 1991,
Mathis 1994).
Therefore, it has been suggested that mantles form on the bump grains in dark clouds 
resulting in broad bumps.  In bright nebulae, there are no mantles and narrower bumps
result from the bare grains.
In this scenario, mantles are able to form in dense clouds shielded
from the interstellar radiation field while the mantles on grains near H~II regions
are removed by the stronger radiation field.
However, the three small $\gamma$ lines of sight in the LMC appear to be associated
with a dense environment even though they are near the 30 Dor star forming region.
Several stars in the LMC--general sample (eg. SK $-$66 19 and SK $-$66 88) are associated
with bright H~II regions and yet have normal Galactic bump widths.
Accepting the explanation
for the narrow bumps based on the Galactic data, we would expect to find
narrow bumps in the LMC~2 sample.  In contrast with this expectation,
the data presented in Table~3 indicates
that the LMC~2 bump widths are comfortably within the average Galactic range.
It doesn't appear that the trend in $\gamma$ with environment seen the Galaxy
holds in the LMC.
There are no exceptionally wide bumps in our sample save
SK $-$70 116 with $\gamma=1.4$; however, the bump is extremely weak and $\gamma$
is not strongly constrained.

The weak bumps in the LMC~2 region are not unique.  As discussed above, several
Galactic lines of sight also have extinction curves with very weak bumps (HD 29647, 
HD 62542, HD 210121).  However, the LMC~2 environment seems to have little in 
common with these Galactic lines of sight which in turn seem to have little in
common with each other.  Though the swept up, shocked environment near HD 62542
may be similar to the LMC~2 environment (but on a vastly reduced scale), the other
two Galactic sightlines sample relatively quiescent environments.  
HD 210121 lies behind a single diffuse, translucent cloud about 150~pc from the 
Galactic plane. The interstellar radiation
field is weaker than in the general interstellar medium and shocks do not appear
to be important (Welty \& Fowler 1992).  Larson et. al. (1996) suggest that the apparent
preponderance of small grains along the HD 210121 line of sight is due to lack of
grain growth through coagulation as a result of lack of time spent in a dense environment.
It appears that
very diverse environmental conditions result in rather similar bump profiles.
It is not known whether 
the bump grains are being modified in a similar fashion
in different environments or substantially different modifications of the bump 
grains can result in a similar UV extinction in the bump.

\section{Conclusions}
Evidently the relationship between the UV extinction, dust grain properties, and
environment is a complicated one. Similar variations
in the form of the UV extinction can arise in a variety of environments.
The environmental dependences
seen in the Galaxy do not seem to hold in the LMC.  
Since large variations in UV extinction are seen within both the LMC and the Galaxy,
global parameters
such as metallicity cannot be directly responsible for the observed variations from
galaxy to galaxy as has been suggested (e.g., Clayton \& Martin 1985).
However, one effect of decreased metallicity in the LMC 
is that the typical molecular cloud is
bigger but more diffuse than those in the Galaxy (Pak et. al. 1998).
Hence, dust grains in the LMC may not spend as much time in dense, shielded environments
as grains in the Galaxy.  The lack of time in dense environments may contribute to
the apparent small size of the LMC grains as indicated by the low values of $R_V$ measured
in this study.
In addition, the weak and narrow bump lines of sight in the
LMC all lie near the 30 Dor star forming region which has no analog in the Galaxy.  
The dust along these sightlines
has probably been affected by the proximity to the harsh environment
of the copious star formation associated with 30 Dor.
However, it must be pointed out that the most extreme UV extinction
curves, having virtually no bumps and a very steep far UV are found in the SMC.
The SMC dust lies near regions of star formation but they are very modest compared 
to 30 Dor. These SMC sightlines have optical depths
similar to those in LMC~2 (GC).
Due to very low metallicity of of the SMC, its molecular clouds are very diffuse (Pak et. al. 1998).
One might expect that values of $R_V$ in the SMC be even smaller than in the LMC; the
current observations, however, show no evidence for this (GC).

Even with the improved and expanded samples of extinction in the LMC and SMC,
the link between particular environments and dust characteristics is still unclear.
The combination of the Galactic and Magellanic cloud data show that the extinction 
curve/environment links are not as simple as previously proposed.  
But the different times spent by grains in dense molecular environments may be a 
significant factor as suggested for the Galactic star HD 210121 (Larson et. al. 1996).
The processing history of dust grains (ie. coagulation and mantling in dense clouds 
environments and exposure to strong shocks and radiation field outside of clouds) 
is probably quite different in these three galaxies owing to the different
molecular cloud environments and the varying intensity of star formation.
The interplay between at least these two factors likely plays an important role in determining
the form the UV extinction.  The fact that starburst galaxies appear to have SMC--type
dust regardless of metallicity (Calzetti et. al. 1994; 
Gordon et. al. 1997) implies that the star formation
history of a galaxy plays an important role in determining the extinction properties. 
However, the complicated relationship between extinction properties in the UV and environment
implied by the Galactic and Magellanic Cloud data suggests that great care must be taken in
assuming the form of the UV extinction in external galaxies.

\acknowledgments
This research has made use of the SIMBAD database.  $IUE$ spectra were down loaded
from the $IUE$ final archive at ESA. 
This work has been partially supported through NASA ATP grant NAG5~3531 to GCC.
We thank M. Oestreicher for providing the source code and data files used for 
generating the foreground reddening map and M. Bessel for supplying the H$\alpha$ image.

\end{document}